\documentclass[useAMS,usenatbib]{mnras}

\usepackage{graphicx}
\usepackage{enumerate}
\usepackage{xcolor} 
\usepackage{amsmath}
\newcommand{\hers}{{\it Herschel}}
\newcommand{\hst}{{\it HST}}
\newcommand{\logoh}{12$+$log(O/H)}

\newcommand{\msun}{$M_\odot$}
\newcommand{\msunyr}{$M_\odot$\,yr$^{-1}$}

\newcommand{\hii}{H{\sc ii}}

\newcommand{\mstar}{M$_{*}$}
\newcommand{\ha}{H$\alpha$}
\newcommand{\hb}{H$\beta$}
\newcommand{\oiii}{[O{\sc iii}]}
\newcommand{\siii}{[S{\sc iii}]}
\newcommand{\oii}{[O{\sc ii}]}
\newcommand{\nii}{[N{\sc ii}]}
\newcommand{\te}{T$_{\rm e}$}

\newcommand{\meann}{$\langle n \rangle$}
\newcommand{\meanb}{$\langle \beta \rangle$}
\newcommand{\ltir}{$L_{\rm TIR}$}
\newcommand{\lha}{$L_{\rm{H}\alpha}$}

\title[Metallicity and SFR coevolution]{Coevolution of metallicity and star formation 
in galaxies to $z\simeq 3.7$: I. A Fundamental Plane }
\author[Leslie Hunt, Pratika Dayal, Laura Magrini, Andrea Ferrara]{Leslie Hunt$^{1}$\thanks{E-mail: hunt@arcetri.astro.it}, Pratika Dayal$^{2}$, 
Laura Magrini$^{1}$ and Andrea Ferrara$^{3}$ \\
$^{1}$INAF/Osservatorio Astrofisico di Arcetri, Largo Enrico Fermi 5, 50125 Firenze, Italy\\
$^{2}$Kapteyn Astronomical Institute, University of Groningen, P.O. Box 800,
9700 AV Groningen, The Netherlands\\
$^{3}$Scuola Normale Superiore, Piazza dei Cavalieri 7, I-56126 Pisa, Italy}
\begin{document}

\date{draft version 10 May 2016}

\pagerange{\pageref{firstpage}--\pageref{lastpage}} \pubyear{2016}

\maketitle

\label{firstpage}

\begin{abstract}
With the aim of understanding the coevolution of star formation rate (SFR),
stellar mass (\mstar), and oxygen abundance (O/H) in galaxies up to redshift
$z\simeq 3.7$, we have compiled the largest available dataset for studying
Metallicity Evolution and Galaxy Assembly (MEGA); it comprises 
$\sim$1000 galaxies with a common O/H calibration and 
spans almost two orders of magnitude in metallicity, 
a factor of $\sim10^6$ in SFR, 
and a factor of $\sim10^5$ in stellar mass. 
From a Principal
Component Analysis, we find that the 3-dimensional parameter space reduces to a
Fundamental Plane of Metallicity (FPZ) given by $12+\log({\rm O/H})  =
-0.14\,\log {\rm (SFR)} + 0.37\,\log ({\rm M_*)} + 4.82$. 
The mean O/H FPZ
residuals are small (0.16\,dex) and consistent with trends found in smaller
galaxy samples with more limited ranges in \mstar, SFR, and O/H.
Importantly, the FPZ is found to be approximately redshift-invariant within the uncertainties. 
In a companion paper, these results are
interpreted with an updated version of the model presented by \citet{dayal13}. 
\end{abstract}


\section{Introduction}
\label{sec:intro}

Galaxies are assembled over cosmic time by the accumulation of stellar mass (\mstar) through
star-formation (SF) processes.
This build-up is accompanied by an increase of metal content, typically measured through
the gas-phase oxygen abundance (O/H), the most abundant heavy element produced by
massive stars.
Stellar mass is a measure of the integrated SF activity over the history of the galaxy, while
the star-formation rate (SFR) indicates the current rate for conversion of gas into stars.
The gas-phase metallicity ($Z$) reflects not only the metal production from high-mass stars,
but also the level of galaxy interactions with environment through inflows and outflows
in the form of galactic winds.

Given the causal relation between star-formation processes and metal content
in galaxies, it is not surprising that \mstar, SFR, and O/H are mutually
correlated. 
The mass-metallicity relation \citep[MZR, e.g.,][]{tremonti04} is a
manifestation of the \mstar\,$- Z$ correlation; the SF ``main sequence'' relates
\mstar\ and SFR \citep[SFMS, e.g.,][]{brinchmann04,salim07,noeske07}.
The mutual relations among the three variables extend to specific SFR (sSFR\,$\equiv$\,SFR/\mstar)
and metallicity which are also correlated \citep[e.g.,][]{salim14,yates14}.

These mutual correlations imply that residuals from the main relations (MZR, SFMS)
should be correlated with the third variable.
Indeed, from an analysis of data from the Sloan Digital Sky Survey (SDSS),
\citet{mannucci10} found an expression that connected the residuals in the
MZR to SFR; this was dubbed the ``Fundamental Metallicity Relation" (FMR) and
reduced the scatter in O/H over $\sim$80\,000 galaxies from $\sim$0.1\,dex
to 0.05$-$0.06\,dex.
In a similar vein, \citet{laralopez10} showed that 
the 3D space of \mstar, SFR, and O/H 
for $\sim$33\,000 SDSS galaxies
could be expressed through a two-dimensional
(planar) surface (``Fundamental Plane", FP).
By fitting regressions to parameter pairs, they expressed the FP in terms of \mstar\
and found a residual variation of $\sim$0.16\,dex, larger however than that
found for the FMR.

Given that reducing a three-dimensional (3D) parameter space to a (2D) plane is mathematically
equivalent to diagonalizing the 3D covariance matrix, a natural approach to this
problem is a Principal Component Analysis (PCA).
A PCA
was first applied to \mstar, SFR, and O/H by \citet{hunt12} for $\sim$1000 galaxies
from $z\sim 0 - 3.5$ selected to span a range of 
$\ga 10^5$ in SFR and two orders of magnitude in O/H\footnote{To avoid
problems with the curvature of the MZR at high metallicities, \mstar\ was limited
to $\leq$10.5\,dex\,\msun, so those results are formally applicable
only to galaxies less massive than this limit.}.
The PCA showed that the principal component dominated by O/H was the
component {\it most dependent} on the other two, 
as by itself it comprised only $\sim$2\% of the total variance. 
The PCA resulted in a FP in metallicity (FPZ) with a spread of 0.17\,dex in O/H,
despite the vast range in the original parameters, including redshift.
This FPZ applied to the same SDSS samples used by \citet[][here mass
limited]{mannucci10} gave roughly the same residuals as the FMR,
0.06\,dex. Thus, \citet{hunt12} concluded that the FPZ could be used to estimate metallicities with
an accuracy of $\sim 40-50$\% over an extended range of \mstar\ and SFR, and moreover was a good representation of O/H at $z\ga3$.

It is now well established that both the MZR and the SFMS extend to the highest redshifts examined so far,
but with differing normalizations relative to the Local Universe;
at a given \mstar, SFR (and sSFR) increases with increasing redshift
\citep[e.g.,][]{noeske07,elbaz11,karim11,wuyts11,speagle14} while metallicity decreases
\citep[e.g.,][]{erb06a,maiolino08,mannucci09,cresci12,xia12,yabe12,henry13,cullen14,zahid14,troncoso14,steidel14,wuyts14,ly15,delosreyes15}.
Consequently, if we assume that the FPZ is redshift-invariant (an assumption
that we shall reassess below), the higher sSFRs found in high$-z$ galaxy populations must
be related, perhaps causally, to the lower metallicities observed at the same redshift. 
This is the hypothesis we examine in this paper.

In order to observationally constrain the evolution of metallicity with
redshift, we have compiled a new dataset of $\sim 1000$ star-forming galaxies 
from $z\simeq 0$ to $z\sim3.7$ with nebular oxygen abundance measurements;
we will refer this compilation as the ``MEGA'' dataset, corresponding to 
{\it Metallicity Evolution and Galaxy Assembly}.
This compilation is a radical improvement over the dataset used by \citet{hunt12}
because of the inclusion of several more high$-z$ samples 
and, more importantly, because of a common metallicity calibration.
Section \ref{sec:samples} describes the 19 individual samples which form
the MEGA compilation, together with our estimates of stellar masses and SFRs
for the samples at $z\simeq 0$.
The procedures for aligning the individual samples to a common O/H calibration
are outlined in Sect. \ref{sec:ohcalib}.
Sect. \ref{sec:scaling} describes the scaling relations for the MEGA dataset 
and
re-evaluates the redshift invariance of the FPZ through
a linear analysis of the correlations of \mstar, SFR, and O/H in the MEGA sample
and in $\sim 80\,000$ galaxies
at $z\sim0$ selected from the SDSS by \citet{mannucci10}. 
The coevolution of SFR and O/H with redshift in the MEGA dataset
is presented in Section \ref{sec:coevo}, together with a comparison of 
results with previous work. 
We discuss our results and summarize our conclusions in Sect. \ref{sec:discussion}.
Throughout the paper we use a \citet{chabrier03} Initial-Mass Function (IMF)
and, when necessary, adopt the conversions for \mstar\ and SFR given by \citet{speagle14}.

\section{Galaxy samples}
\label{sec:samples}

Because of the need to compare stellar mass, \mstar,
SFR, and metal abundance [as defined by the
nebular oxygen abundance, \logoh], we have selected only samples of galaxies 
for which either these quantities are already available in the
literature, or can be derived from published data.
Here we discuss the estimates of \mstar\ and SFR;
the metallicity determinations for the samples will be discussed
in Sect. \ref{sec:ohcalib}.

\subsection{Local Universe}
Four samples of galaxies in the Local Universe met these criteria:
the 11\,Mpc distance-limited sample of nearby galaxies or
Local Volume Legacy \citep[11HUGS, LVL:][]{kennicutt08,lee09,lee11};
the Key Insights into Nearby Galaxies: a Far-Infrared Survey with \hers\ \citep[KINGFISH,][]{kennicutt11};
the starburst sample studied by \citet{engelbracht08}, and 
the blue compact dwarf (BCD) sample by \citet{hunt10}.
There are 15 galaxies that appear both in the KINGFISH and LVL samples;
for these, we used the KINGFISH parameters from \citet{kennicutt11} because
of the uniform O/H calibration given by \citet{moustakas10}.
The starbursts from \citet{engelbracht08} were restricted
to only those galaxies (42) with 
metallicities derived from the ``direct'' or ``\te" method based on electron temperatures
(see Table \ref{tab:samples} and Sect. \ref{sec:ohcalib});
the BCDs (23) all have \te-measured metallicities.

\subsubsection{Star-formation rates }
\label{sec:sfr}

In order to maximize consistency, 
we have recalculated SFRs and \mstar\ for the four local samples
starting from photometric fluxes reported in the literature.
SFRs were derived according to \citet{murphy11} using
for KINGFISH and LVL the hybrid method with far-ultraviolet (FUV)$+$total infrared luminosity (\ltir);
these data were available for 123 (of 138 non-KINGFISH) galaxies with O/H in the LVL
and for 50 (of 55) KINGFISH galaxies.
For the KINGFISH and LVL galaxies without these data, we adopted other SFR calibrations
given by \citet{murphy11} including TIR (5 galaxies in KINGFISH, 6 LVL),
UV (3 LVL), and \ha$+$24\,\micron\ (4 LVL).
\ltir\ was calculated according to \citet{draine07} and fluxes
were taken from \citet{dale07,dale09} and \citet{lee09,lee11}.
For 20 LVL galaxies, the SFRs inferred from \ltir\ were larger than those from
FUV$+$\ltir\ using the prescriptions by \citet{murphy11};
in those cases, we adopted SFR(\ltir).
For NGC\,253 and M\,82, SFR(\ltir) is $\sim$2 times SFR(FUV$+$\ltir),
but for the other galaxies the two estimates agree to within 30\%.
We also compared for the LVL galaxies 
the SFRs calculated with FUV$+$\ltir\ with those inferred
by combining \ha\ and 24-\micron\ luminosities \citep[\lha, $L_{24}$:][]{calzetti10,murphy11};
SFR(FUV$+$\ltir) tends to be
$\sim$1.6 times larger than SFR(\ha$+L_{24}$) with a scatter of $\sim$0.2\,dex.
This is consistent with the findings of \citet{leroy12} who found
a similar trend at low surface SFR densities such as those in the LVL galaxies.

Because FUV data are generally not available for the starbursts or the BCDs, 
for these we adopted
the hybrid combination of \ha$+L_{24}$\,\micron\ as prescribed by \citet{murphy11}.
Total \ha\ fluxes were taken from
\citet{dopita02,gildepaz03,james04,pustilnik04,cannon05,moustakas06,schmitt06,lopezsanchez08,kennicutt08,cairos10,james10}
and 24\,\micron\ measurements from \citet{engelbracht08}.
When these data were unavailable (9 galaxies), we adopted the SFR(TIR) prescription
by \citet{murphy11} using the fluxes by \citet{engelbracht08}. 
For SBS\,0335$-$052 and I\,Zw\,18, we adopted the SFRs from
radio free-free emission \citep{hunt04,hunt05,johnson09}, given
the superiority of such estimates over other methods \citep[e.g.,][]{murphy11}.
For one galaxy in the starburst sample (UM\,420), because of the lack of MIPS observations, the SFR was estimated
from \ha\ luminosities, and for one galaxy (ESO\,489$-$G56) there were no data available from which to infer SFR
(so it was not considered further).
For the (23) BCDs from \citet{hunt10}, total \ha\ fluxes were taken from \citet{gildepaz03,rosagonzalez07,perez11,lagos14},
and 24\,\micron\ fluxes from Hunt et al. (2016, in prep.).
As for the starbursts, for the three galaxies without \ha\ data, we used SFR(TIR); and
for SBS\,1030$+$583 there were no MIPS data so we adopted SFR(\ha).

Considering the different SFR estimators discussed above,
and considering their varying degrees of applicability, for the local samples
the uncertainties on the SFRs are probably around a factor of 2 (0.3\,dex).
As mentioned above, the SFRs have been reported to a \citet{chabrier03} IMF.

\subsubsection{Stellar masses}
\label{sec:mstar}

We calculated the stellar masses according to \citet{wen13},
a method 
based on WISE W1 (3.4\,\micron) luminosities.
This approach exploits the approximately constant mass-to-light ratios
of stellar populations at near-infrared wavelengths, 
independently of metallicity and age \citep{norris14,mcgaugh14}.
However, when W1 photometry was not available, we used
IRAC 3.6\,\micron\ photometry instead.
In fact, the two bands are very similar;
using data from \citet{brown14}, \citet{grossi15} find for spirals a 
mean flux ratio $F_{3.4}/F_{3.6}\,=\,1.02\pm0.035$.
Including also the data for dwarf irregulars from \citet{brown14} we find a
mean flux ratio $F_{3.4}/F_{3.6}\,=\,0.98\pm0.061$.
Thus, we conclude that the ratio of the W1 and IRAC 3.6\,\micron\ bands is
unity, with 5-6\% scatter for galaxies like our targets.

For the starburst and BCD samples, we used the \hii-galaxy formulation by \citet{wen13}, rather than what
they found for their full sample; the \hii\ galaxies have the lowest mass-to-light
ratios in their compilation, corresponding roughly to the bluest regions of
the galaxies studied by \citet{zibetti09}. 
To better take into account the weak trends with abundance found by \citet{wen13}, 
we also applied an approximate correction 
for low metallicity \citep[by multiplying the mass-to-light ratio
by 0.8 when \logoh$\leq$8.2:][see their Fig. 17]{wen13}.
Instead, for the LVL and KINGFISH samples, we adopted the 
\citet{wen13} formulation based on morphological type, and considered
an ``early-type'' galaxy one with Hubble type $T<2$\footnote{The distinction used by \citet{wen13}
is based on colors which are not available for all our samples.}.

However, before applying the relations by \citet{wen13}, we first
subtracted nebular emission and emission from hot dust where possible.
In starbursts and BCDs,
such contamination can be very important in the near-infrared
and can contribute 50\% or more to the observed flux at these wavelengths
\citep{hunt01,hunt02,smith09,hunt12}.
The ionized gas continuum contribution to the 3.4-3.6\,\micron\ flux was 
estimated from the SFR using the emission coefficients from \citet{osterbrock06}.
When possible, we also subtracted the hot-dust component, with
the assumption that $H$-band emission is entirely stellar.
Because $H$-band photometry is available for some of our sample, we used the data
from \citet{brown14} to estimate the maximum possible IRAC 3.6\,\micron/$H$-band
ratio in galaxies similar to our targets; 
95\% of the spiral/dwarf irregular galaxies have a flux ratio $\leq$2.4.
This corresponds to a (Vega-based) [H-3.6] color of $\sim0.8$, consistent
with what is found for the pure stellar component in star-forming galaxies
\citep{hunt02}.
After subtraction of the nebular component, any excess over this ratio was
attributed to hot dust and subtracted; this subtraction was not possible
for 33 galaxies, including all the BCDs.

We compared the stellar masses obtained with the formulation of 
\citet{wen13} to those calculated according to 
\citet{lee06} based on IRAC 4.5\,\micron\ luminosities \citep[used by][]{hunt12}.
For the BCDs, the masses based on \citet{wen13} are on average $-$0.2\,dex lower
than those based on \citet{lee06} with a scatter of 0.15\,dex;
this is not unexpected given the blue colors of these galaxies
and the results of \citet{zibetti09} who showed that the \citet{bell01}
calibration used by \citet{lee06} gives mass-to-light ratios
that are too high for such blue galaxies.
Instead for the starbursts the two estimates are in closer agreement,
with $-$0.06\,dex difference on average and a scatter of 0.20\,dex.
Moreover, for the galaxies having both W1 and IRAC data, the stellar masses
obtained from 3.6\,\micron\ luminosities are within $\sim$5\% of those
from W1 as expected.

For LVL, we have compared our estimates with those from \citet{cook14}
who used a constant mass-to-light ratio and the IRAC 3.6\,\micron\ luminosity.
The Wen-derived stellar masses are on average $-$0.25\,dex smaller than
the \citet{cook14} values, with a scatter of 0.09\,dex. 
\citet{skibba11} derived stellar masses for the KINGFISH sample according
to the formulation of \citet{zibetti09} based on optical and $H$-band colors.
We have compared ours derived using \citet{wen13} to theirs and find the values
from \citet{skibba11} are smaller by $\sim$0.5\,dex on average, with a 0.3\,dex scatter.

Given the significant uncertainties inherent in the procedures to
derive stellar masses over a wide range of galaxy types,
and considering the offsets and scatters of our new \mstar\
estimates, 
the uncertainties on the stellar masses for the local samples
are at 
most 
a factor of 2 (0.3\,dex). 
As above, these values are based on a \citet{chabrier03} IMF.

\subsection{SDSS10 $z\simeq 0$ galaxy sample}

\citet{mannucci10} analyzed a set of emission-line galaxies from the SDSS,
using the stellar masses from \citet{kauffmann03}, and SFRs measured
from \ha\ after correcting for extinction using the Balmer decrement;
they reported all values to a \citet{chabrier03} IMF.
The parameter range covered by this sample is much more limited
than the MEGA sample:
9.2\,$\la$\,dex(\mstar)$\la$\,11.3\,\msun;
8.5\,$\la$\,\logoh$\la$\,9.1 (assuming the \citealt{kewley02} calibration, see below);
$-1.3\,\la$\,log(SFR)$\la$\,0.8\,\msunyr.
Nevertheless, we include this sample, hereafter SDSS10, in our analysis because of its 
superb statistics for 
comparison both locally and at $z>0$. 

\begin{figure}
\vspace{\baselineskip}
\includegraphics[height=0.4\textwidth]{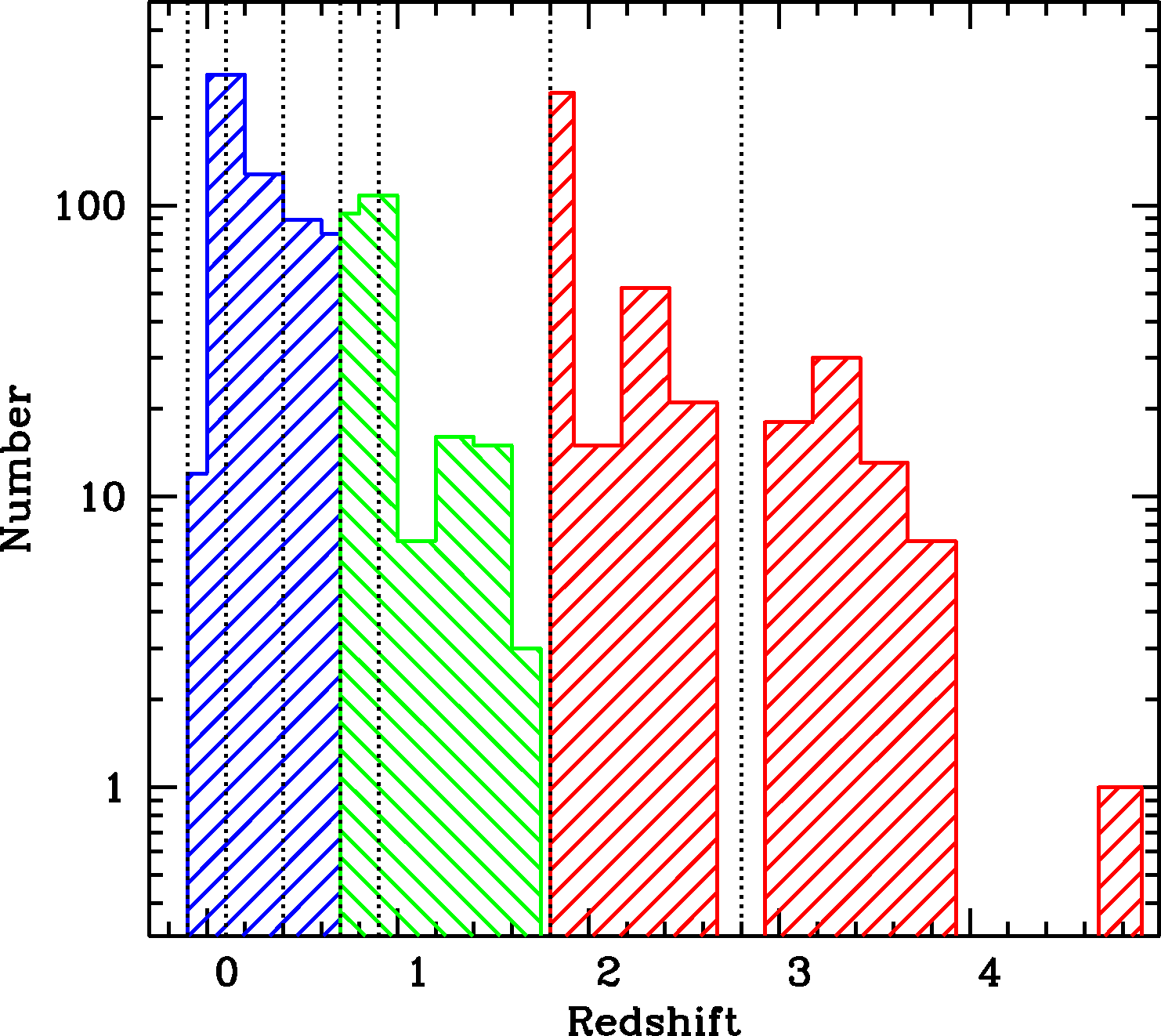}
\caption{Redshift distribution of combined sample (without SDSS10).
The 7 redshift bins used throughout the paper for the MEGA dataset
are also shown as vertical dotted lines.
The colors of the portions of the histogram are arbitrary, with the aim
of illustrating ``low" redshift (blue), ``intermediate" redshifts (green),
and ``high" redshifts (red).
}
\label{fig:redshift}
\end{figure}

\begin{table*} 
\caption{Characteristics of the individual samples in the MEGA dataset}
\begin{tabular}{lcrccll}
\hline 
\multicolumn{1}{c}{Parent} &
\multicolumn{1}{c}{Redshift} &
\multicolumn{1}{c}{Number}  &
\multicolumn{1}{c}{Selection} &
\multicolumn{1}{c}{Original} &
\multicolumn{1}{c}{SF method} &
\multicolumn{1}{c}{Reference} \\
\multicolumn{1}{c}{sample} &
\multicolumn{1}{c}{range} &&
\multicolumn{1}{c}{criterion} &
\multicolumn{1}{c}{O/H calibration} \\
\hline 
\multicolumn{7}{c}{Local Universe} \\
KINGFISH   & $-0.001-0.008$ &  55 & Representative    & KK04   & FUV$+$TIR$^{\mathrm a}$ & \citet{kennicutt11} \\
LVL        & $-0.001-0.003$ & 138 & Volume-limited    & M91, KK04, Direct$^{\mathrm b}$ & FUV$+$TIR$^{\mathrm a}$ & \citet{kennicutt08} \\
Starburst  & $   0.0-0.058$ &  41 & Representative    & Direct$^{\mathrm c}$ & \ha$+$24\micron$^{\mathrm d}$ & \citet{engelbracht08} \\
BCD        & $ 0.009-0.044$ &  23 & Primordial helium & Direct$^{\mathrm c}$ & \ha$+$24\micron$^{\mathrm d}$ & \citet{hunt10} \\
\\
\hline 
\multicolumn{7}{c}{$0.1 \leq z \la 0.9$} \\
COSMOS     & $ 0.17-0.91$ & 334 & $I$ band                         & KD02   &  \ha, \hb   & \citet{cresci12} \\
COSMOS     & $ 0.62-0.69$ &  26 & \oiii\,$\lambda\lambda$4959,5007 & KK04   & SED fitting & \citet{henry13} \\
DEEP2      & $ 0.71-0.91$ &  27 & \oiii\,$\lambda$4363             & Direct &  \hb        & \citet{ly15} \\
New\ha\    & $ 0.79-0.82$ & 143$^{\mathrm e}$ & Narrow-band \ha\   & T04    &  \ha        & \citet{delosreyes15} \\
\hst-grism & $0.60-2.32$  &  11 & \oiii\,$\lambda\lambda$4959,5007, \oii\,$\lambda$3727 & KK04 &  \hb  & \citet{xia12} \\
\\
\hline 
\multicolumn{7}{c}{$z > 0.9$} \\
DEEP2      & $1.02-1.40$  &  9  & $R$ band                         & PP04N2   &   \ha      & \citet{shapley05} \\
DEEP2      & $1.02-1.40$  &  7  & $R$ band                         & PP04N2   &   \ha      & \citet{liu08} \\
VVDS       & $1.27-1.53$  &  6  & \oii\,$\lambda$3727              & T04      &   \ha      & \citet{queyrel09} \\
BX         & $2.11-2.43$  &  7  & $U_n-G$, $G-{\cal R}$ colors     & PP04N2   &   \ha      & \citet{shapley04} \\
KBSS       & $2.02-2.55$  & 79  & $H$ band$^{\mathrm f}$ & PP04O3N2 & \ha$^{\mathrm g}$  & \citet{steidel14} \\
LSD        & $2.93-3.41$  &  8  & Lyman-break dropout              & KD02     &   \ha      & \citet{mannucci09} \\
AMAZE      & $3.04-4.87$  & 26  & Lyman-break dropout              & KD02     &   \ha      & \citet{troncoso14} \\
COSMOS     & $2.97-3.69$  & 35  & Predicted \hb\                   & KD02     &   UV       & \citet{onodera16} \\
\\
\hline 
\multicolumn{7}{c}{Stacked samples} \\
SXDS/UDS   & $1.27-1.52$  & 5$^{\mathrm h}$ & $K$ band            & PP04N2 &   \ha      & \citet{yabe14} \\
COSMOS     & $1.40-1.70$  & 10$^{\mathrm h}$ & sBzK                & PP04N2 &   \ha      & \citet{zahid14} \\
\\
\hline
\label{tab:samples} 
\end{tabular} 
\vspace{-\baselineskip}
\begin{flushleft}
$^{\mathrm a}$ ~If not available, then SFR(FUV), or as last choice SFR(TIR). \\ 
$^{\mathrm b}$ ~Taken from \cite{berg12} or \citet{marble10} when available, otherwise
from \citet{moustakas10} (KK04). \\
$^{\mathrm c}$ ~Taken from 
\citet{berg12},
\citet{guseva03a},
\citet{guseva03b},
\citet{guseva11},
\citet{guseva12},
\citet{izotov04},
\citet{izotov06},
\citet{izotov07},
\citet{izotov09},
\citet{izotov12},
\citet{kobulnicky96},
\citet{kobulnicky97},
\citet{kniazev03},
\citet{kniazev04},
\citet{mattsson11},
\citet{perez05},
\citet{roennback95},
\citet{shi05},
\citet{thuan05},
\citet{vigroux87},
\citet{zhao10}. \\
$^{\mathrm d}$ ~If not available, then the maximum of SFR(TIR) and SFR(\ha). \\
$^{\mathrm e}$ ~AGN have been excluded.\\
$^{\mathrm f}$ ~This is only one of several ``layered'' criteria for selecting the galaxies for KBSS Keck-MOSFIRE observations.\\
$^{\mathrm g}$ ~SFRs are taken from \citet{steidel14a}. \\
$^{\mathrm h}$ ~These are from stacked spectra, but are treated here as individual measurements;
the redshifts are taken as the average given in the respective papers 
\citep[$z\sim1.4$, and $z\sim1.6$, for][respectively]{yabe14,zahid14}.\\
\end{flushleft}
\end{table*}

\subsection{$z>0$ samples}

Because our analysis is focused on observationally constraining metal
content at high redshift, to construct the MEGA dataset
we have culled from the literature all available
samples at $z>0$ with measured \mstar, SFR, and O/H.
Stacked analyses have been avoided
where possible, and are used only to increase statistics when 
tabulations of observations for individual galaxies were not available
in the required redshift range. 
We identified 14 samples at $z>0.1$ (see Table \ref{tab:samples}) for which
these three parameters were measured.
Unavoidably, this compilation is subject to a variety of selection
effects which change with sample and redshift.
Nevertheless, from the observational point of view, the MEGA dataset
constitutes a unique tool with which to assess basic
trends among \mstar, SFR, and O/H, and establish how they vary with
redshift.
Table \ref{tab:samples} lists the samples that comprise the MEGA
dataset, together with their redshift range, selection technique,
and other information.
We postpone the important discussion of metallicity estimates to
Sect. \ref{sec:ohcalib}.

\subsubsection{$0.1 \leq z \leq 0.9$}

The most important representative samples in the redshift range $0.1\leq\,z\,\la\,0.9$ come from two surveys,
zCOSMOS \citep{lilly09,cresci12} and New\ha\ \citep{delosreyes15}; these two
datasets alone comprise 477 galaxies.
The first, from COSMOS, is $I$-band selected and was first described by \citet{lilly09}.
Stellar masses were derived from fitting spectral energy distributions
(SEDs) of 12 photometric bands, including {\it Spitzer}/IRAC data
at 3.6$-$5.8\,\micron.
SFRs were calculated from \ha\ and \hb\ luminosities, after correcting
for extinction either via the Balmer decrement (for galaxies with $z\la0.49$)
or using the extinction estimated from the SED fitting with an appropriate
multiplicative factor.
The second large sample in this redshift range comes from the New\ha\ survey, 
selected from narrow-\ha\ band images designed to identify emission-line
galaxies around $z\approx 0.8$.
\citet{delosreyes15} calculate stellar masses through
SED fitting of eight photometric bands (up to observed frame $J$ band),
and estimate SFRs from
the \ha\ images after correcting for the contribution from \nii\
and for extinction.

Unlike \citet{hunt12}, we do not include in the MEGA dataset the
Luminous Compact Galaxies (LCGs) by \citet{izotov11} and the ``Green Peas''
\citep{amorin10}; the latter galaxies are selected by bright \oiii\,$\lambda$5007
emission in the SDSS $r$ band \citep{cardamone09}. 
LCGs, instead, are defined by requiring an \oiii\,$\lambda$4363
detection, large \hb\ equivalent width (EW), and a flux limit in \hb. 
Thus the LCGs are young (because of the high \hb\ EW), highly star forming (because of the \hb\ flux limit)
and metal poor (because of the \oiii\,$\lambda$4363 detection).
Although highly interesting objects, the \citet{izotov11} selection criteria
favor young, metal-poor galaxies, and thus are not be representative of abundances
of typical galaxy populations at those redshifts.

There are $\sim$60 galaxies in the remaining three samples in this redshift range:
galaxies selected from a multi-slit narrowband spectroscopic survey with
\oiii\,$\lambda\lambda$4959,5007, \oii\,$\lambda \lambda$3727,3729 at $z\sim0.6-0.7$ by \citet{henry13};
\oiii\,$\lambda$4363 DEEP2 selected objects at $z\sim0.7-0.9$ by \citet{ly15};
and galaxies selected from {\it HST}-grism observations (\oiii, \oii) by \citet{xia12}.
These three samples are very interesting because of their selection methods
which tend to favor less massive galaxies than typical broadband photometry
selections.
Stellar masses were derived from SED fitting of COSMOS imaging data including IRAC
bands \citep{henry13}; of 8-band photometry up to $z^\prime$ for the DEEP2
survey \citep{ly15}; and
of 10-band {\it HST} ACS/WFC3 photometry up to F160W \citep{xia12}.
With the exception of \citet{henry13} who used SED fitting to calculate SFRs,
\ha\ and \hb\ corrected luminosities were used to infer SFRs.

\subsubsection{$z > 0.9$}

Most of the galaxies in this redshift range are color-selected
Lyman-Break Galaxies \citep[e.g.,][]{steidel99}.
However, the \citet{queyrel09} galaxies are
selected from the magnitude-limited Mass Assembly Survey with SINFONI
in VVDS \citep[MASSIV,][]{epinat09},
and the two $z\sim1$ samples 
by \citet{shapley05} and \citet{liu08} are selected from
the DEEP2 Galaxy Redshift Survey \citep{davis03}. 
Wavelength coverage for stellar-mass determinations varies,
with $U_n G {\cal R} K_s$ \citep{shapley04};
$BRI K_s$ \citep{shapley05,liu08};
$UBVRIZ_s JK$ \citep{queyrel09};
and 14 spectral bands from GOODS-MUSIC \citep{grazian06},
including IRAC 3.6, 4.5\,\micron\ \citep{mannucci09,troncoso14}.
Stellar masses for the \citet{onodera16} COSMOS sample 
are fit with $uBVrizYJHK$ and IRAC bands. 
\citet{onodera16} 
prefer SFRs inferred from extinction-corrected UV luminosities,
but all other SFRs in this redshift range
are determined from \ha\ suitably corrected for extinction.

To ensure better coverage of the redshift range $1.3 < z < 1.7$,
we have included also the two samples by 
\citet{yabe14}
and \citet{zahid14}.
Neither group publishes data for individual galaxies, 
so we have
adopted the parameters of their stacked spectra here 
as individual galaxies, and used the average redshifts
of $z\sim1.4$ and $z\sim1.6$ for \citet{yabe14,zahid14}, respectively.

The redshift distribution of the MEGA dataset is shown in Fig. \ref{fig:redshift},
together with the 7 redshift bins that will be used throughout the paper.
As mentioned above,
Table \ref{tab:samples} gives the characteristics of the 19 individual samples comprising
the MEGA dataset; there is a total of 990 galaxies from $z\simeq 0$ to
$z \sim 3.7$ (and LnA1689$-$2 in the AMAZE sample at $z\,=\,4.87$).

\section{Metallicity calibrations}
\label{sec:ohcalib}

Oxygen abundance O/H is typically used as a proxy for metallicity in
emission-line galaxies.
Because the ionized gas in \hii\ regions at lower metal abundance is hotter (as measured by electron temperature, \te), 
the preferred technique to establish O/H is to measure \te\ 
and the physical conditions in the ionized plasma.
In this ``direct-temperature" or ``\te" method, the \te\ 
of the ionized gas is derived from the ratio of the \oiii\,$\lambda$4363 auroral line to lower-excitation lines
(\oiii\,$\lambda$4959,\,5007); 
such flux ratios are sensitive to temperature because the auroral and strong lines
originate from different excitation states (second and first excited states, respectively).
Because the oxygen transitions are collisionally excited, the relative population
of the excited states depends on \te.
Thus, the strengths of these forbidden lines,
combined with the measurement of \te\ and density in the nebula, can be converted to an 
abundance, relative to hydrogen, 
after 
correcting for unseen phases of ionization \citep[e.g.,][]{osterbrock06}.

Although the \te\ method is more directly related to metallicity,
the auroral lines are weak and often difficult to detect, especially at high metallicity.
Thus, ``strong-line" methods are more generally used to estimate O/H,
especially in metal-rich objects and at high redshift. 
It is necessary to calibrate these methods, either using
theoretical photoionization models \citep[e.g.,][hereafter KD02]{kewley02},
or measurements of \te\ \citep[e.g.,][hereafter PP04]{pettini04},
or a combination of the two
\citep[e.g.,][hereafter D02]{denicolo02}.
Despite the best efforts to correctly cross calibrate these methods over
a wide range of physical conditions,
there remain large discrepancies, as high as 0.6\,dex in log(O/H)
\citep[e.g.,][and references therein]{kewley08}.
Thus to correctly assess metal content and its evolution with redshift,
it is necessary to apply a common metallicity calibration to the
samples under discussion. 

In nearby galaxies where spectra can be obtained with sufficient signal-to-noise,
the \te\ method is generally used.
As mentioned above,
the most widely used auroral \te\ diagnostic line is \oiii\,$\lambda$4363 because of 
its relative ease of observation, high abundance of emitting ions,
and notable strength in the low- and intermediate-metallicity regime (i.e., below solar metallicity).
However, there are several potential problems with the \te\ method based on \oiii:
\begin{enumerate}[{\em i)}]
\item
Metallicities derived from collisionally-excited lines (CELs) such as \oiii\ can be underestimated when temperature fluctuations inside the nebula are present but neglected. The assumption of a single average CEL temperature for the whole nebula, usually higher than the temperature derived from the Balmer discontinuity, tends to lead to an underestimate of the abundances \citep[e.g.,][]{peimbert67,stasinska05,bresolin07,perez10,pena12}.
\item
Additional problems also plague the \te\ method including
possible non-Boltzmann electron distributions \citep[e.g.,][]{nicholls12,binette12,nicholls13};
depletion of oxygen onto dust grains \citep[e.g.,][]{peimbert10,pena12};
and potential shock waves within the nebulae \citep[e.g.,][]{binette12}.
\item
Finally, recent results suggest that metallicities derived from \oiii\ may
be more unreliable than those from other auroral lines such as \siii\,$\lambda$6312
and \nii\,$\lambda$5755 \citep[e.g.,][]{berg15}; however, these lines are even more
difficult than \oiii\ to measure in distant galaxies.
\end{enumerate}
An alternative to the use of \te-diagnostic lines can be the derivation of abundances from optical recombination lines (ORLs), 
because of their reduced emissivity dependence on density and temperature.
Abundances derived from the ratio of the intensity of ORLs tend to be systematically higher than those from CELs  
\citep[e.g.][]{peimbert93,liu95,liu01,tsamis04,ge07}. However, such differences may arise from the relation of the ORL abundances to 
small H-deficient portions of the regions, while the CEL-based metallicities are more representative of the whole nebula \citep[see, e.g.,][]{liu00}.
Moreover, such lines are extremely faint, thus requiring very
high signal-to-noise spectra that are currently available only for the Galaxy
and the Local Group \citep[e.g.,][]{blanc15}.

There are also, perhaps more severe, problems with ``strong-line" methods,
and the simplifying assumptions made for photoionization model calibrations
\citep[e.g., photoionization structure, geometry, stellar age: see][for a thorough discussion]{moustakas10}.
As for the \te\ method,
there may also be systematic discrepancies due to the metallicity-dependent correction
for the depletion of oxygen onto dust grains \citep[e.g.,][]{peimbert10}.
Ultimately, the \te\ method (with \oiii) is generally considered to be the most viable, given
the limitations with other techniques.

Thus, to ensure the best possible comparison among different
samples that rely on different O/H calibrations,
it is advantageous to use the strong-line calibration method that
most closely resembles values inferred from oxygen-based \te-method estimations.
According to the results of \citet{andrews13},
who used a stacking technique to measure the oxygen abundances of
$\sim$200\,000 star-forming galaxies from the SDSS to enhance the 
signal-to-noise ratio of the weak \oiii\,$\lambda$4363 line,
there are three such methods:
PP04 (both \nii\ and \oiii$+$\nii-based: hereafter PP04N2, PP04O3N2)
and D02.
Over the metallicity and \mstar\ range covered by their calculations
of various strong-line methods \citep[see Fig. 10 of][]{andrews13},
the discrepancies between the \te\ method and these three methods
are $\la$0.1\,dex in \logoh.

Thus, in what follows, where there are no direct-\te\ estimates,
we have applied the transformations
given by \citet{kewley08} to convert the original strong-line O/H
calibrations for the MEGA dataset (and SDSS10 sample) 
to the calibrations by D02 and PP04 (PP04N2, PP04O3N2).
As reported in Table \ref{tab:samples},
the original O/H calibrations include:
KD02 \citep[][]{kewley02,cresci12,mannucci10,troncoso14}; 
KK04 \citep[][]{kobulnicky04,kennicutt11,henry13,xia12};
M91 \citep[][]{mcgaugh91,marble10};
PP04N2, PP04O3N2 \citep[][]{pettini04,shapley04,shapley05,liu08,yabe12,zahid14,steidel14};
and T04 \citep{tremonti04,delosreyes15}.

\begin{figure*}
\hbox{
\includegraphics[width=\textwidth]{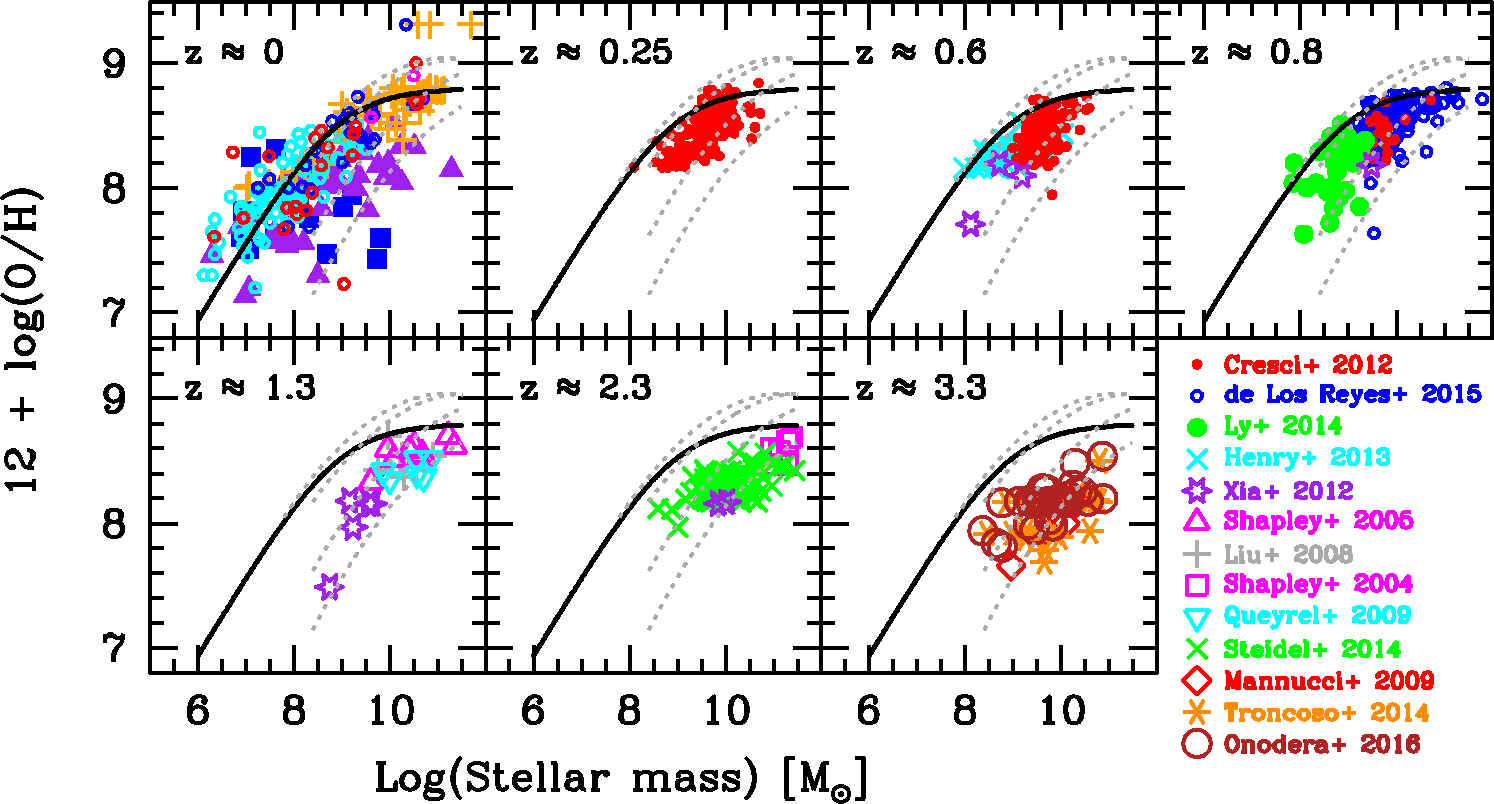}
}
\caption{Mass-metallicity relation over redshifts from $z\sim0$ to $z\ga3.3$,
binned as described in the text.
The solid (black) curve (labeled $z\sim0$) corresponds to the fit to SDSS10
(with stacked \te\ metallicity determinations) by
\citet{andrews13}, and the dotted (grey) curves to the polynomial
fits by \citet{maiolino08} with the KD02 calibration at
$z \approx 0.07$, $z \approx 0.7$, $z \approx 2.2$, and $z \approx 3.5$.
The O/H calibration for all galaxies is PP04N2 as described in the text.
Samples are labeled according to the legend in the lower rightmost panel,
except for $z\ \approx\ 0$ which are: LVL as small open circles 
(colors correspond to Hubble types with late types (T$\geq$8) as cyan,
5$\leq$T$<$8 as blue, 3$\leq$T$<$5 as magenta, T$<$3 as red);
KINGFISH as (orange) $+$;
\citet{engelbracht08} as (purple) filled triangles;
\citet{hunt10} as (blue) filled squares.
The stacked samples \citep{yabe14,zahid14} at $z \approx 1$ are not plotted.
}
\label{fig:mzr}
\end{figure*}

\begin{figure*}
\hbox{
\includegraphics[width=\textwidth]{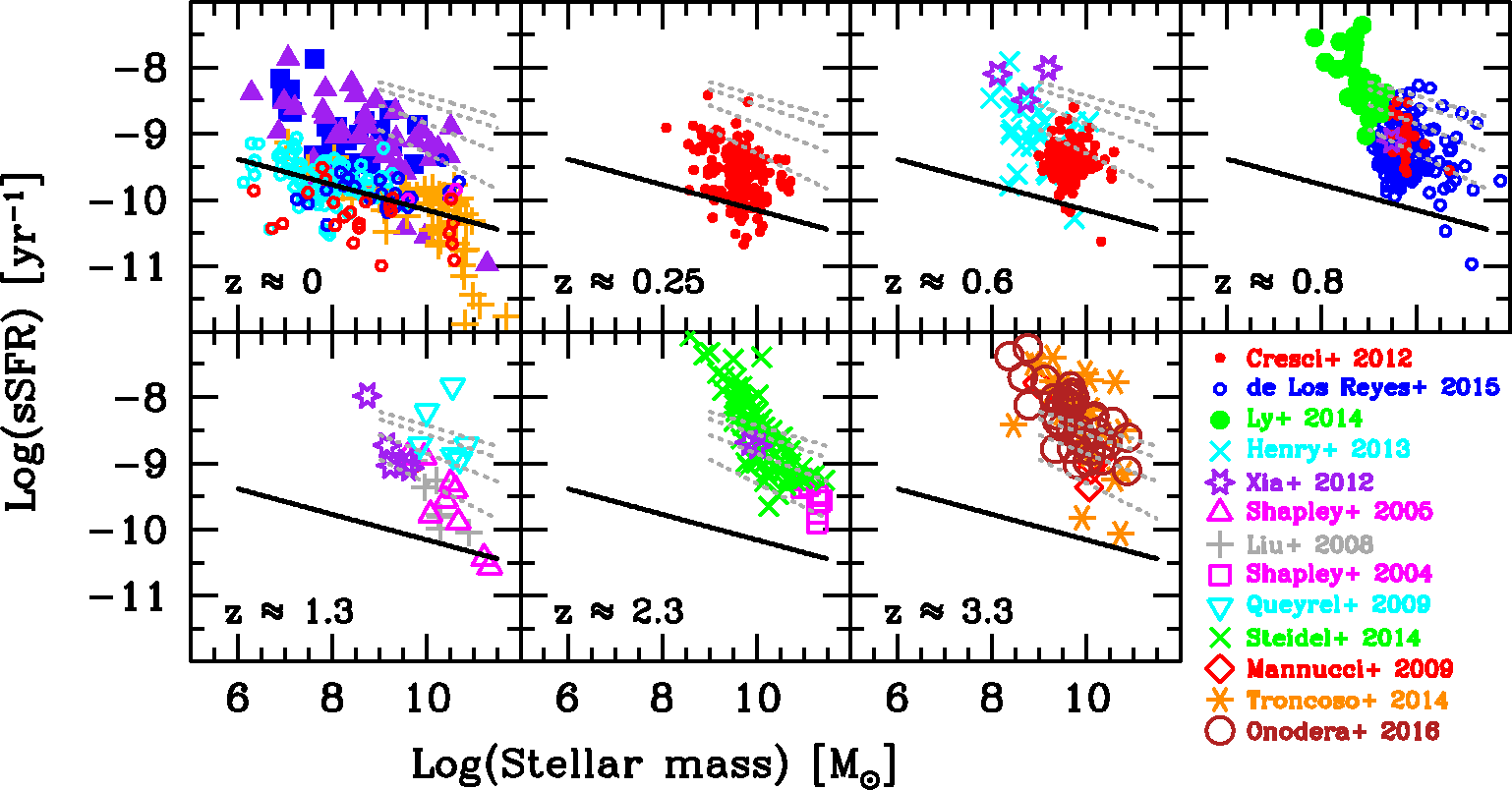}
}
\caption{Specific SFR vs. \mstar\ (main sequence of star formation) 
over redshifts from $z\sim0$ to $z\ga3.3$, binned as described in the text.
The solid (black) curve (labeled $z\sim0$) corresponds to the fit to the LVL$+$KINGFISH
samples, and
the dotted (grey) curves to the formulation of dependence with
\mstar\ and $z$ by \citet{speagle14} for $z\approx 0.6$,
$z\approx 1.3$, $z\approx 2.3$, and $z\approx 3.3$.
Symbols are as in Fig. \ref{fig:mzr},
and the stacked samples \citep{yabe14,zahid14} are not plotted.
}
\label{fig:ms}
\end{figure*}

\section{Scaling relations and the fundamental plane}
\label{sec:scaling}

The MEGA dataset comprises
three parameters (pseudo-observables, as they are not directly observed):
nebular oxygen abundance (\logoh), stellar mass (\mstar), and SFR. 
As discussed in the Introduction, these three parameters are mutually correlated,
although O/H trends flatten at high \mstar\ (and high O/H).
Here we discuss the scaling relations of the three parameters:
the mass-metallicity relation, MZR,
the ``main sequence'' of star formation, SFMS,
and the correlation (at least at $z\sim0$) between sSFR and metallicity.

\begin{figure*}
\hbox{
 \includegraphics[width=0.95\textwidth]{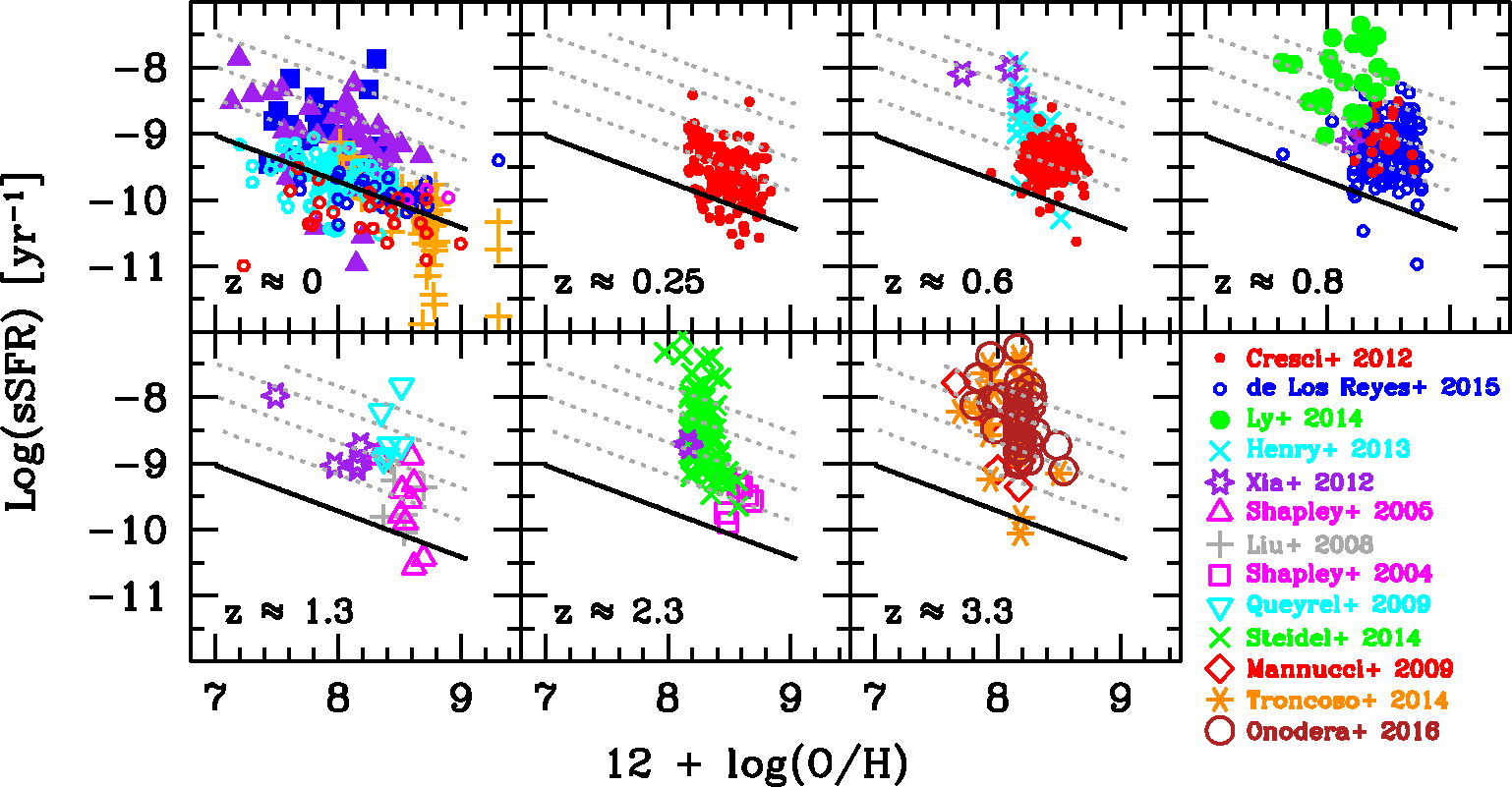}
}
\caption{Specific SFR vs. \logoh\ 
over redshifts from $z\sim0$ to $z\ga3.3$, binned as described in the text.
The solid (black) line 
corresponds to the fit to the best-fit slope for the MEGA sample at $z\sim0$,
and
the dotted (grey) lines to the prediction of the redshift variation
assuming the increase in sSFR given by \citet{speagle14}
(from $z\sim0.6$, $z\sim1.3$, $z\sim2.3$, $z\sim3.3$) 
of the Fundamental Plane for these two parameters described in Sect. \ref{sec:fp}.
As in Fig. \ref{fig:mzr},
the O/H calibration for all galaxies is PP04N2 as described in the text.
Symbols are as in Fig. \ref{fig:mzr},
and the stacked samples \citep{yabe14,zahid14} are not plotted.
}
\label{fig:ssfroh}
\end{figure*}

The MZR with the PP04N2 O/H calibration
for different redshift bins is shown in
Fig. \ref{fig:mzr}. 
The solid curve shows the \te-method MZR derived by \citet{andrews13}
which well approximates the MEGA dataset at $z \approx 0$.
The dotted grey curves represent the polynomial fits given by \citet{maiolino08}
for the KD02 calibration;
at high \mstar, these curves fail to capture the \te-derived (or PP04N2) metallicities
because of the different O/H calibration.
As virtually all previous work suggests,
the different panels illustrate that as $z$ increases, at a given \mstar\
metallicity decreases.
However, at $z \approx 0$ for a given \mstar,
the starburst and BCD samples tend to be more metal-poor than the LVL and KINGFISH galaxies;
they behave more like galaxies at $z \ga 1$ than like galaxies in the Local Universe,
presumably because of their higher sSFR.

The high sSFRs in the starburst and BCD $z \approx 0$ samples are more clearly seen in
Figure \ref{fig:ms}, which shows the SFMS, or sSFR plotted against \mstar.
The solid line shows the SFMS calibrated with the LVL$+$KINGFISH samples,
having a slope of $-0.19\,\pm\,0.02$, roughly consistent with that ($-0.23$) found by \citet{elbaz07} for $z\sim0$ galaxies.
The dashed grey lines correspond to the \citet{speagle14} formulation for SFR as a function
of cosmic time (we have calculated cosmic age for representative redshifts and plotted the result).
The slope by \citet{speagle14} at $z\ga2$ is similar to what we find for the Local Universe,
which however is shallower (steeper in SFR-\mstar\ space)
than their value for $z\sim0$. 
Fig. \ref{fig:ms} illustrates that as redshift increases, for a given \mstar,
sSFR also increases; galaxies that would be main-sequence galaxies at $z\ga1$ are starbursts if
found at $z\approx0$.
However, it is also seen from the figure that the individual high-$z$
samples do not clearly 
follow the SFMS;
this is almost certainly due to selection effects and will be further discussed
in Sect. \ref{sec:massvariation}. 
Because of the difficulty in measuring metallicities in high-$z$ emission-line galaxies,
flux limits for spectroscopy impose a commensurate limit in SFRs.

\begin{figure*}
\vspace{\baselineskip}
\hbox{
 \includegraphics[width=0.95\textwidth]{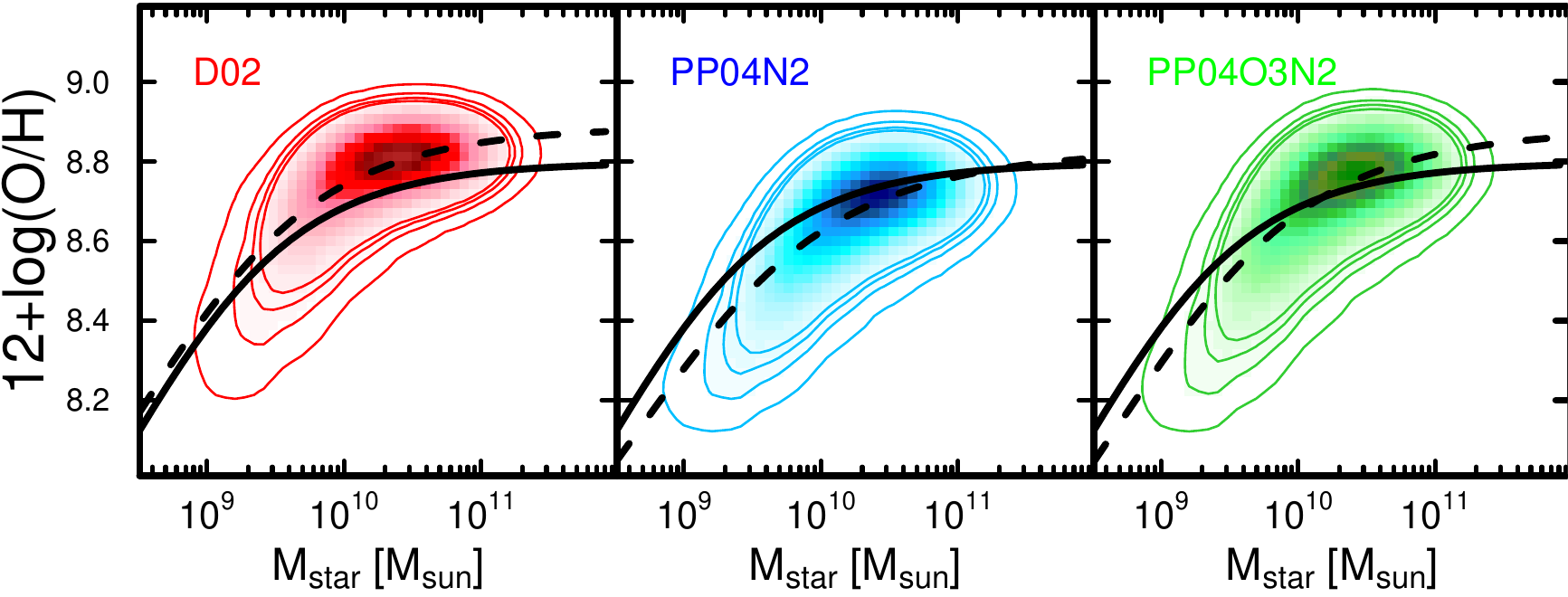}
}
\caption{SDSS10 galaxies: \logoh\ plotted against (log of) \mstar. 
The three panels correspond to number densities of
\logoh\ (with the three O/H calibrations) and \mstar.
In each panel, the solid curves 
correspond to the mass-metallicity relation taken from \citet{andrews13} 
for stacked O/H direct-temperature metallicity determinations as also shown
in Fig. \ref{fig:mzr}.
The dashed curves show instead the 
relation of the form used by \citet{andrews13} but fit to the SDSS10 data used here. 
}
\label{fig:mzrsdss}
\end{figure*}

The third correlation between sSFR and O/H is shown in Figure \ref{fig:ssfroh}.
As in previous figures, the solid line gives the local calibration on the LVL$+$KINGFISH 
and the dotted grey lines show the redshift trend for O/H expected for the higher SFR
as predicted by the FPZ (see Sect. \ref{sec:fp}).
The SFRs of the local starbursts and BCDs are higher at a given O/H, relative to the
other local samples; again, they are more similar to galaxies at $z\ga1$ than to typical
local populations.
Similarly to the behavior of the MEGA dataset for the SFMS, the LVL$+$KINGFISH galaxies 
show a well-defined correlation between sSFR and O/H, but the correlation disappears
for the higher-redshift samples.

The main point of this third correlation is that, at least locally, the three
psuedo-observables, \logoh, SFR, and \mstar\ are mutually interdependent.
This makes it difficult to determine which is the primary parameter(s) driving
the relations, and it this point which we explore below in
Sect. \ref{sec:fp}.

\subsection{The SDSS10 relations}
\label{sec:sdss}

Similar correlations are found for the SDSS10 galaxies, although
the range in \mstar, O/H, and sSFR is smaller than in the MEGA dataset.
Nevertheless, over the limited parameter range the sheer number statistics
afford precise determinations of scaling relations and fitting functions
which will be important for constraining our models.

Fig. \ref{fig:mzrsdss} gives the MZR for the three O/H calibrations of SDSS10 sample,
transformed from the original KD02.
The solid curves, also shown in Fig. \ref{fig:mzr},
give the MZR for the direct-method O/H as found by \citet{andrews13},
while the dashed ones are functions of the same form but fit to
the SDSS10 dataset itself.
The PP04 calibrations (middle and right panels) are, on average, the best approximation to the
direct-method O/H curve, although at \logoh\,$\la$8.5, the D02 calibration
is superior.
In any case, the functional MZR form used by \citet{andrews13} does not well approximate the 
SDSS10 data at low mass or low metallicities.
The low-mass, low-metallicity linear portion of the data has a slope of $\sim 0.38\,\pm\,0.003$,
similar to the MZR curve, but the latter is offset to higher masses.

\begin{figure}
\vspace{\baselineskip}
\hbox{
 \includegraphics[width=0.45\textwidth]{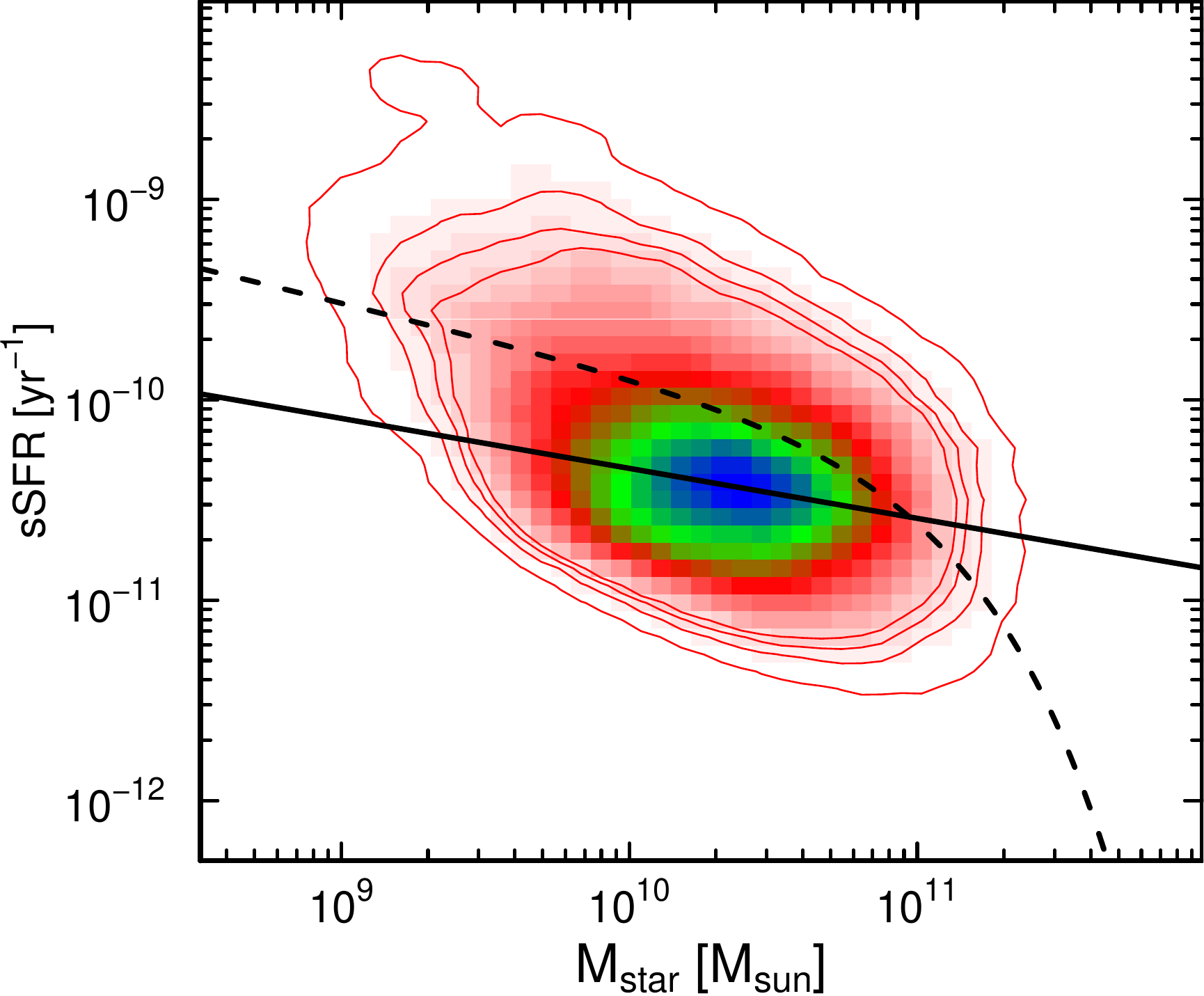}
}
\caption{SDSS10 galaxies: sSFR plotted against \mstar.
The color scale corresponds to the number densities of the two parameters.
The solid curve gives the linear SF ``main sequence'' regression
to the LVL$+$KINGFISH samples (also shown in Fig. \ref{fig:ms}), and
the dashed curve to the Schechter function fitted by \citet{salim07}.
}
\label{fig:mssdss}
\end{figure}

The SFMS of the SDSS10 data is shown in Fig. \ref{fig:mssdss}; the solid line corresponds
to the linear regression for the LVL$+$KINGFISH galaxies (also shown in Fig. \ref{fig:ms}) and
the dashed curve to the Schechter-like functional form fitted by \citet{salim07}.
Although this last captures the low- and high-mass ends of the SDSS10 data, it does
not pass through the region with the highest density (blue colors in Fig. \ref{fig:mssdss}).
This could have something to do with the different ways that SFR is calculated;
\citet{salim07} used FUV while \citet{mannucci10} used extinction-corrected \ha.
Nevertheless, the LVL$+$KINGFISH regression well approximates this behavior, implying
that the SFR derivation is probably not the cause of the discrepancy.
The third correlation, between sSFR and \logoh\ is not shown for the SDSS10 data;
it shows a similar behavior to the MEGA sample.

\subsection{A planar approximation to scaling relations}
\label{sec:fp}

At high \mstar\ and O/H, both the MZR and SFMS inflect and flatten
\citep[e.g.,][]{tremonti04,noeske07,whitaker14,lee15,gavazzi15}.
However, for \mstar\ below a certain threshold, \mstar$\leq3\times10^{10}$\,\msun, roughly the ``turn-over mass''
\citep{tremonti04,wyder07}, the relations among the variables
are approximately linear. 
We propose that, at high \mstar\ and O/H, the inflections the MZR and the SFMS compensate one another,
and hypothesize that even above this inflection threshold, 
the trends in \mstar, O/H, and SFR can be approximated by linear relations.
Consequently, as discussed above,
these observationally-defined variables could
define a {\it plane} which, given the relatively large scatters in the SFMS, the
MZR, and the SFR-O/H relation, is not viewed in the best projection.
Because the three parameters are mutually correlated, it is important to determine which
of the three is the most fundamental, and whether or not the planar
approximation is sufficient to describe the data.
This can be readily accomplished through a Principal Component Analysis
\citep[PCA, e.g.,][]{hunt12}.

The MEGA dataset is a significant improvement on the sample studied by \citet{hunt12},
and is particularly well suited for such an analysis. 
In particular, the MEGA dataset 
triples the number of galaxies at $z\ga2-3$ with respect to \citet{hunt12}. 
It spans
almost two orders of magnitude in metallicity (\logoh\,=\,7.1 to $\sim9$),
a factor of $\sim10^6$ in SFR ($\sim10^{-4} \leq$ SFR $\leq\, \sim10^2$\,\msunyr),
and a factor of $\sim10^5$ in stellar mass ($\sim10^6 \leq$ \mstar\ $\leq\, \sim10^{11}$\,\msun);
moreover it includes galaxies at redshifts from $z \sim 0-3.8$ (see Fig. \ref{fig:redshift}).
Other samples previously analyzed to find scaling relations cover much smaller
parameter ranges: typically less than a decade in metallicity (\logoh$\geq$8.4), 
a factor of $\sim$200 in SFR ($\sim$0.04$\la$ SFR$\la 6$\,\msunyr),
and roughly 2 orders of magnitude in stellar mass (\mstar$\ga 10^9$\,\msun)
\citep[e.g.,][]{tremonti04,mannucci10,laralopez10,yates12}. 
Because more than 50\% of the MEGA dataset has $z > 0.5$,
and it includes galaxies at redshift $z \ga 3.5$,
we can test the assumption that the relations among the observationally-defined variables
are redshift invariant.

\begin{figure*}
\vspace{\baselineskip}
\includegraphics[width=\textwidth]{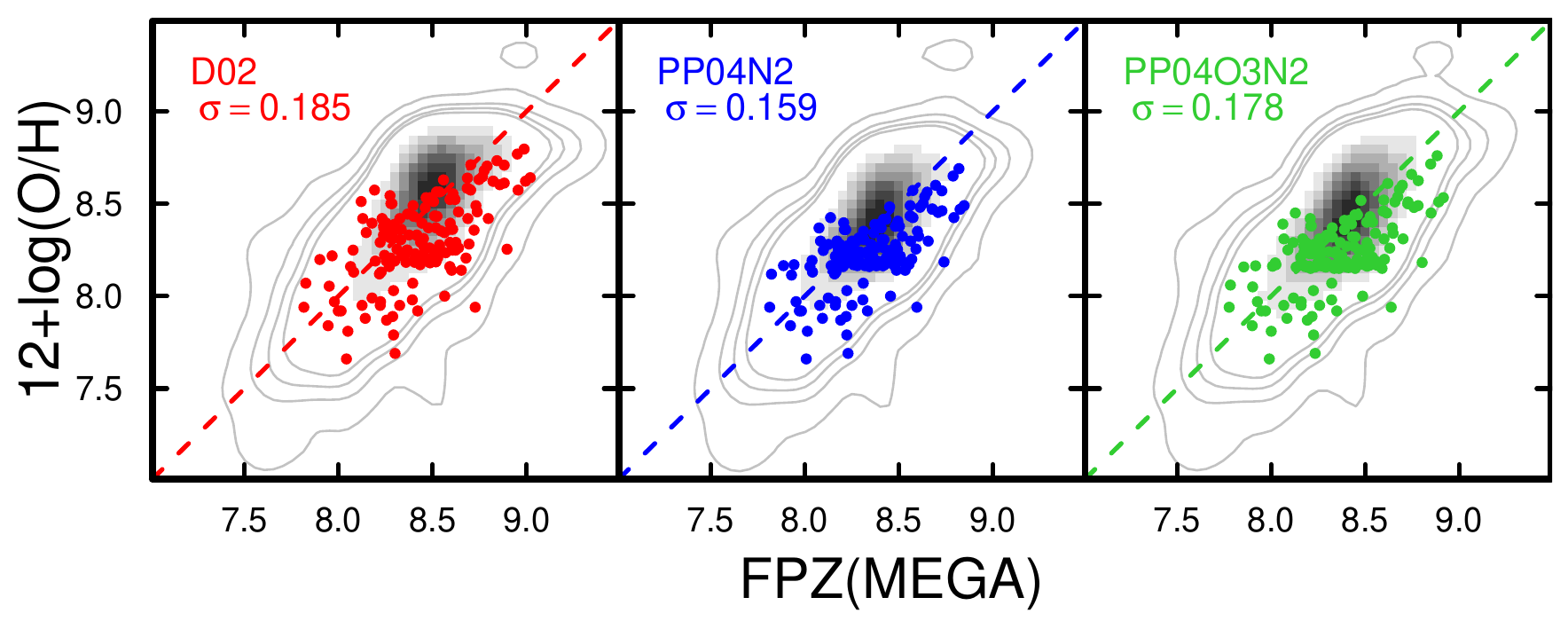}
\vspace{\baselineskip}
\includegraphics[width=\textwidth]{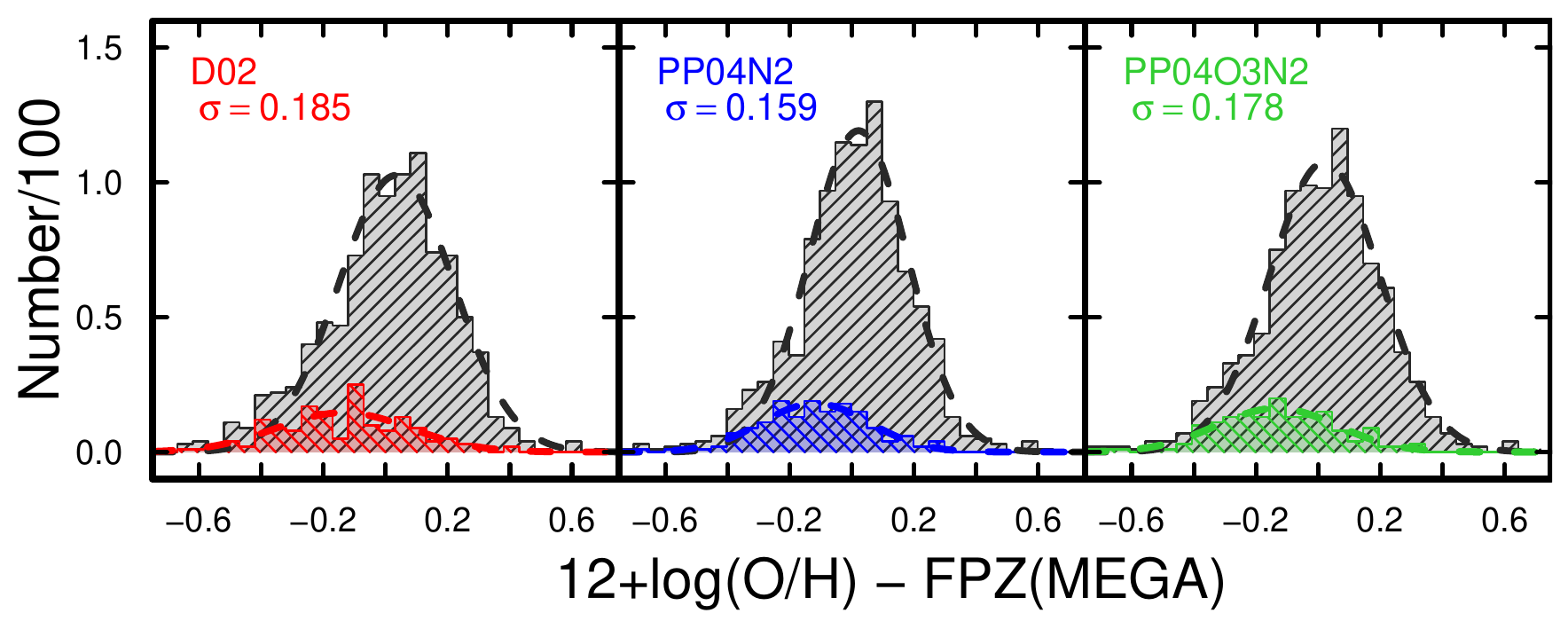}
\caption{Top panel: FPZ projection of O/H for all O/H calibrations
of the MEGA dataset as described in the text.
All galaxies are shown (in gray intensity and contours), including those with \mstar $\geq 3\times10^{10}$\,\msun;
colored filled circles correspond to galaxies with $z\geq2$.
The dashed lines give the identity relation.
Bottom panel: histograms of FPZ residuals from the identity relation.
The grey histograms show all galaxies, while the (lower-amplitude) colored ones correspond
to galaxies with $z\geq2$.
The heavy dashed lines give the Gaussian fit of the residuals
(including the separate fit to the galaxies with $z\geq2$); 
the $\sigma$ values in the upper left corner of each panel correspond to the mean residuals
for all galaxies.
The width of the best (PP04N2) residual distribution over all redshifts is $\sim$0.16\,dex (see middle panels).
}

\label{fig:fpzall}
\end{figure*}

\subsection{PCA of the MEGA and SDSS10 samples}
\label{sec:pca}

We have therefore performed a PCA for all three O/H calibrations
of the MEGA dataset 
without imposing a limit in \mstar\ \citep[c.f.,][]{hunt12}.
A PCA diagonalizes the 3D covariance matrix, 
thus defining the orientation of the parameter space which minimizes the covariance.
The orientation is contained in the eigenvectors which are,
by definition, mutually orthogonal.
If the 3D space formed by the three psuedo-observables is truly planar, 
we would expect most of the variance to be contained in the first 
two eigenvectors (the orientation of the plane); 
for the third eigenvector, perpendicular to the plane, the variance should be 
very small.

Independently of the O/H calibration,
the PCA shows that the set of three observables truly defines a plane;
$\ga$98\% of the total variance is contained in the first two eigenvectors.
Most (87\%) of the variance is contained
in the first eigenvector alone (or Principal Component, PC), PC1;
it is dominated by SFR, with \mstar\ contributing slightly
less, and O/H giving only a marginal contribution.
PC2, the second eigenvector, holds 10$-$11\% of the variance, 
and is dominated by \mstar, followed by SFR, and as in PC1,
with O/H again only marginal.
The smallest fraction of the variance ($\sim 1.5-1.8$\%) is contained in the third
eigenvector, PC3, which is dominated by O/H;
the implication is that O/H is the most {\it dependent} parameter,
governed almost completely by \mstar\ and SFR.
Moreover, this means that the 3D space defined by O/H, SFR, and \mstar\ is degenerate; 
because of the mutual correlations of the psuedo-observables,
only two parameters are required to describe the properties of the galaxies. 

We have also performed a PCA of the SDSS10 galaxies, and
obtained similar results: namely, the third eigenvector, PC3, the one
dominated by O/H, contains the smallest fraction of the variance.
The residuals of 0.05$-$0.06\,dex are comparable to that obtained
by the FMR formulation by \citet{mannucci10}.

That star-forming galaxies form a plane in O/H, SFR, and \mstar\ 
is not a new result. 
\citet{laralopez10} concluded that the 3D space of O/H, SFR, and \mstar\ of
$\sim$ 33\,000 SDSS galaxies could be represented as a plane
but used a regression analysis rather than a PCA
\citep[although see][]{laralopez13}. 
\citet{hunt12} derived a PCA for a dataset similar to ours, although dominated
by LCGs, and also concluded that a 2D plane was sufficient to describe the 3D dataset.

As in the Introduction,
we will refer to the resulting 2D plane as the FPZ (Fundamental
Plane in metallicity).
The FPZ for the MEGA dataset
is shown in the top panel of Figure \ref{fig:fpzall} where we have plotted \logoh\ vs.
the equation that results from equating PC3 (PP04N2) to zero (see Table \ref{tab:fpzfmr}
for the other O/H calibrations):

\begin{equation}
12+\log(O/H) = -0.14\,{\rm log(SFR)} + 0.37\,{\rm log(M_*)} + 4.82 \ \ 
\label{eqn:fpzall}
\end{equation}

\noindent
The bottom panel of Fig. \ref{fig:fpzall} shows the residuals from the FPZ
for the different O/H calibrations.
For PP04N2, they are well approximated by a Gaussian with a $\sigma\,=\,0.16$,
corresponding to $\la$45\% uncertainty; the
other two O/H calibrations (D02, PP04O3N2) give similar results,
although slightly larger (see Table \ref{tab:fpzfmr}).
The residuals of the FPZ relation are independent of redshift to within 0.16\,dex,
the overall uncertainty; 
nevertheless the different symbols plotted in the top panel 
(and the different histograms in the bottom one) suggest some slight deviation with redshift
which we will explore in Sect. \ref{sec:fpzinvariance}.

Because of the turnover of the MZR and SFMS at high stellar masses,
our assumption of linearity in the FPZ could also produce
a residual correlation with \mstar.
We have investigated this possibility and found that the FPZ residuals 
and (log)\mstar\ are uncorrelated; 
the mean residuals of the regression are $\sim$0.16\,dex (for the PP04 calibration),
the same as those of the FPZ itself.
Moreover, the slope of the FPZ residuals vs. (log)\mstar\ is zero to within
the uncertainties ($-0.015\,\pm\,0.01$).
Thus, our hypothesis that the inflections in the MZR and SFMS compensate
one another is apparently justified; the curvature in the MZR can be adequately accommodated by
the increasing SFRs at high \mstar, at least to within the uncertainties of our data.

The FPZ dispersion of $\sim0.16-0.18$\,dex for the MEGA dataset is higher than that found by \citet{tremonti04} for the MZR
defined by 53\,000 galaxies from the SDSS (0.1\,dex), and also higher than the FMR (0.06\,dex)
found for the SDSS10 sample by \citet{mannucci10}. 
However, as shown in Table \ref{tab:fpzfmr},
the SDSS10 data span limited ranges in O/H, SFR, stellar mass (and redshift) relative
to the parameter space covered by the MEGA data.
The mean and standard deviation of (log) stellar mass for the SDSS10 sample is 
$10.26\pm0.41$\,\msun, while the comparable mean, standard deviation for MEGA is $9.44\pm0.93$\,\msun;
(log) SFR shows a similar pattern:
$-0.04\pm0.43$\,\msunyr\ for SDSS10 compared to $0.16\pm1.12$\,\msunyr\ 
for MEGA. 
Thus,
the higher dispersion in the MEGA FPZ is not surprising despite the many more galaxies in SDSS.

The value of the FPZ dispersion is {\it lower} than the scatter of the MZR for $\sim$20\,000 VVDS galaxies
within individual redshift bins from $z \sim 0.3-0.9$ \citep[$\sim$0.20\,dex,][]{lamareille09}. 
The FPZ dispersion for the $\sim$1000 galaxies studied here is only slightly higher than
that found for the MZR of 25 nearby dwarf galaxies \citep[0.12\,dex,][]{lee06}, 
a sample dominated by low-mass galaxies.
It is also only slightly higher than the rms scatter of 0.12\,dex found by \citet{henry13} for
18 galaxies at $z \sim 0.6$ in the mass range dex(8.5)$\leq$\mstar$\leq$dex(9.0). 
Because the dispersion in the MZR is found to increase with decreasing \mstar\
\citep{tremonti04,mannucci11}, 
a $\sigma$ of 0.16\,dex is a reasonable value,
given the broad parameter space covered by our dataset.

\subsection{Comparison with the FMR}\label{sec:compfmr}

\begin{table*} 
\setlength{\tabcolsep}{3pt}
\begin{center} 
\caption {FPZ and FMR applied to MEGA and SDSS datasets$^{\mathrm a}$}
{\scriptsize
\begin{tabular}{lcccccccp{0.15\textwidth}}
\hline 
\multicolumn{1}{c}{Sample} &
\multicolumn{1}{c}{Calibration} &
\multicolumn{1}{c}{$\sigma_{\rm fit}^{\mathrm b}$} &
\multicolumn{1}{c}{Offset(fit)$^{\mathrm b}$} &
\multicolumn{1}{c}{$\langle$\logoh$\rangle$$^{\mathrm c}$} &
\multicolumn{1}{c}{$\langle$Log(SFR)$\rangle$$^{\mathrm c}$} &
\multicolumn{1}{c}{$\langle$Log(\mstar)$\rangle$$^{\mathrm c}$} &
\multicolumn{2}{c}{\logoh\,= } \\
\multicolumn{1}{c}{(1)} &
\multicolumn{1}{c}{(2)} &
\multicolumn{1}{c}{(3)} &
\multicolumn{1}{c}{(4)} &
\multicolumn{1}{c}{(5)} &
\multicolumn{1}{c}{(6)} &
\multicolumn{1}{c}{(7)} &
\multicolumn{2}{c}{(8)} \\
\hline 
\multicolumn{9}{c}{MEGA FPZ applied to the MEGA and SDSS10 datasets} \\
\hline 
\\
MEGA      & D02      & 0.185 & 0.03 & $8.422\,\pm\,0.33$ & $0.159\pm1.12$    & $9.440\,\pm\,0.93$ & 
\multicolumn{2}{c}{$-0.17 \ s +  0.44 \ m +  4.33$} \\ 
MEGA      & PP04N2   & 0.159 & 0.02 & $8.332\,\pm\,0.30$ & $0.159\pm1.12$    & $9.440\,\pm\,0.93$ & 
\multicolumn{2}{c}{$-0.14 \ s +  0.37 \ m +  4.82$} \\ 
MEGA      & PP04O3N2 & 0.178 & 0.03 & $8.355\,\pm\,0.32$ & $0.159\pm1.12$    & $9.440\,\pm\,0.93$ & 
\multicolumn{2}{c}{$-0.16 \ s +  0.41 \ m +  4.48$} \\ 
\\
\hline 
\\
SDSS10    & D02      & 0.102 & $-0.04$ & $8.768\,\pm\,0.09$ & $-0.038\pm0.43$ & $10.263\,\pm\,0.41$ & 
\multicolumn{2}{c}{As above for D02.} \\ 
SDSS10    & PP04N2   & 0.080 & $-0.004$ & $8.660\,\pm\,0.11$ & $-0.038\pm0.43$ & $10.263\,\pm\,0.41$ & 
\multicolumn{2}{c}{As above for PP04N2.} \\ 
SDSS10    & PP04O3N2 & 0.088 & $-0.01$ & $8.709\,\pm\,0.12$ & $-0.038\pm0.43$ & $10.263\,\pm\,0.41$ & 
\multicolumn{2}{c}{As above for PP04O3N2.} \\ 
\\
\hline
\multicolumn{9}{c}{FMR$^{\mathrm d}$ applied to the MEGA and SDSS10 datasets} \\
&&&&&&& \multicolumn{2}{c}{\logoh\,=} \\
&&&&&&&
\multicolumn{1}{c}{$\mu_{0.32}<$9.5\,\msun$^{\mathrm e}$} &
\multicolumn{1}{c}{$\mu_{0.32}\geq$9.5\,\msun$^{\mathrm e}$} \\
\multicolumn{1}{c}{(1)} &
\multicolumn{1}{c}{(2)} &
\multicolumn{1}{c}{(3)} &
\multicolumn{1}{c}{(4)} &
\multicolumn{1}{c}{(5)} &
\multicolumn{1}{c}{(6)} &
\multicolumn{1}{c}{(7)} &
\multicolumn{1}{c}{(8)} &
\multicolumn{1}{c}{(9)} \\
\hline
\\
MEGA      & D02      & 0.175 & $-0.14$ & $8.422\,\pm\,0.33$ & $0.159\pm1.12$    & $9.440\,\pm\,0.93$ & 
$-0.16 \ s + 0.51 \ m + 3.83 $ & $-13.8 + 4.17 m - 1.34 s + 0.12 m\,s - 0.19 m^2  - 0.054 s^2$  \\ 
MEGA      & PP04N2   & 0.168 & $-0.25$ & $8.332\,\pm\,0.30$ & $0.159\pm1.12$    & $9.440\,\pm\,0.93$ & 
$-0.16 \ s + 0.51 \ m + 3.83 $ & $-13.8 + 4.17 m - 1.34 s + 0.12 m\,s - 0.19 m^2  - 0.054 s^2$  \\ 
MEGA      & PP04O3N2 & 0.181 & $-0.22$ & $8.355\,\pm\,0.32$ & $0.159\pm1.12$    & $9.440\,\pm\,0.93$ & 
$-0.16 \ s + 0.51 \ m + 3.83 $ & $-13.8 + 4.17 m - 1.34 s + 0.12 m\,s - 0.19 m^2  - 0.054 s^2$  \\ 
\\
\hline 
&&&&&&& \multicolumn{2}{c}{\logoh\,=} \\
&&&&&&&
\multicolumn{1}{c}{$\mu_{0.32}<$10.2\,\msun$^{\mathrm e}$} &
\multicolumn{1}{c}{$\mu_{0.32}>$10.5\,\msun$^{\mathrm e}$} \\
\multicolumn{1}{c}{(1)} &
\multicolumn{1}{c}{(2)} &
\multicolumn{1}{c}{(3)} &
\multicolumn{1}{c}{(4)} &
\multicolumn{1}{c}{(5)} &
\multicolumn{1}{c}{(6)} &
\multicolumn{1}{c}{(7)} &
\multicolumn{1}{c}{(8)} &
\multicolumn{1}{c}{(9)} \\
\hline
\\
SDSS10    & D02      & 0.05 & $-0.25$ & $8.768\,\pm\,0.09$ & $-0.038\pm0.43$ & $10.263\,\pm\,0.41$ & 
$-0.15 \ s +  0.47 \ m +  4.20$ & \multicolumn{1}{c}{$9.07$} \\
SDSS10    & PP04N2   & 0.05 & $-0.34$ & $8.660\,\pm\,0.11$ & $-0.038\pm0.43$ & $10.263\,\pm\,0.41$ & 
$-0.15 \ s +  0.47 \ m +  4.20$ & \multicolumn{1}{c}{$9.07$} \\
SDSS10    & PP04O3N2 & 0.06 & $-0.28$ & $8.709\,\pm\,0.12$ & $-0.038\pm0.43$ & $10.263\,\pm\,0.41$ & 
$-0.15 \ s +  0.47 \ m +  4.20$ & \multicolumn{1}{c}{$9.07$} \\
\\
\hline 
\label{tab:fpzfmr} 
\end{tabular} 
}
\vspace{-1.5\baselineskip}
\begin{flushleft}
$^{\mathrm a}$~Applying the FPZ and the FMR to the entire mass range according to Eqn. \ref{eqn:fpzall}.
In the equations of Cols. (8,9), $s$ corresponds to Log(SFR) and $m$ to Log(\mstar).  
The sense of the residuals is \logoh(data) $-$ \logoh(FPZ,FMR).\\
$^{\mathrm b}$~Standard deviation $\sigma$ and offset of the FPZ or FMR residuals (e.g., Fig. \ref{fig:fpzall}).\\
$^{\mathrm c}$~Means and standard deviations of the samples. \\ 
$^{\mathrm d}$~We took the FMR for the MEGA sample from the extension to lower \mstar\ by \citet{mannucci11},
and for SDSS10 from \citet{mannucci10}; for both we have converted their recipes with $\mu_{0.32}$ to
the multiplicative formulation as for the FPZ. \\
$^{\mathrm e}$~The divisions for the FMR are in $\mu_{0.32}$, 
where $\mu_{0.32}\,\equiv\,$Log(\mstar) $-$ 0.32\ Log(SFR) \citep{mannucci10}. \\ 
\end{flushleft}
\end{center}
\end{table*}

Table \ref{tab:fpzfmr} gives the mean residuals and offsets of the FPZ applied to the MEGA and SDSS10 datasets,
as well as of the FMR from \citet{mannucci10} applied to SDSS10, and the FMR extended to lower \mstar\
by \citet{mannucci11} applied to the MEGA dataset. 
Despite the vastly different parameter ranges over which the FPZ and FMR are 
calibrated,
results from the Table show that the FPZ and the FMR are roughly equivalent in terms of
the width of the residuals, i.e., the accuracy of the approximation.
The FPZ fits the SDSS10 dataset almost as well as the FMR itself ($\sigma\,\approx\,$0.08\,dex),
and the FMR is reasonably good at reproducing the metallicities of the MEGA dataset.

However, the salient difference between the FPZ and FMR formulations is the negative O/H offsets
of the FMR; Col. (4) of Table \ref{tab:fpzfmr} shows that the FMR predicts metallicities
both for the MEGA dataset and for the recalibrated SDSS10 that can be in excess by as much as $\sim -0.3$\,dex. 
A similar result was found by \citet{hunt12} relative to the FMR,
although the offset was larger, $\sim -0.4$\,dex, presumably because of the original SDSS10 KD02 calibration.

Indeed, the over-large metallicities predicted by the FMR are almost certainly due to the different O/H 
calibrations as discussed in Sect. \ref{sec:ohcalib}.
\citet{cullen14} found a similar discrepancy of their observations at $z \sim 2$
with respect to the FMR using the same (KD02) calibration as \citet{mannucci10}.
Some groups concluded that the FMR evolves with redshift because of
its failure to fit galaxies at $z \sim 2-3$ \citep[e.g.,][]{steidel14,troncoso14}.
However, we find that the FPZ, unlike the FMR, is apparently invariant with redshift;
our result almost certainly stems from the common O/H calibration and its
similarity to the \te\ method by which strong-line methods are calibrated
at low metallicity.
Thus, it is of extreme importance to compare galaxies at different redshifts
with a common O/H calibration that is as accurate as possible at low metallicities,
and that smoothly connects these with the difficult intermediate-metallicity regime 
and with higher metallicities nearer to or exceeding Solar. 

\section{Metallicity and SFR coevolution}
\label{sec:coevo}

Much work has been done to establish how metal abundance and SFR vary with
redshift.
As mentioned in the Introduction,
the picture that emerges from these studies is that
the shape of the MZR is relatively invariant while the metallicity
for a given \mstar\ decreases with increasing redshift
\citep[e.g.,][]{shapley05,cowie08,henry13,yabe14,erb06a,zahid12,steidel14,maiolino08,mannucci09,troncoso14,onodera16}.
At the same time,
it is well known that the SFMS also remains relatively constant in shape,
but at a given \mstar, SFR (and sSFR) increases with redshift \citep[e.g.,][]{noeske07,karim11,speagle14}.
In the context of the FPZ, the relatively small dispersion of the residuals suggests that
the FPZ formulation is apparently invariant with redshift to
$z\sim$3.7, even with the new MEGA sample that more than
triples the number of galaxies at $z\ga2-3$ with respect to \citet{hunt12}.
\textit{Thus, under the hypothesis that the FPZ is maintained even at high $z$,
the opposing redshift trends of O/H and SFR must somehow be mutually compensated. }
For typical galaxy populations, at fixed \mstar,
the increase of SFR with redshift must be accompanied by a corresponding decrease in O/H.
We can quantify such trends with the MEGA dataset and the FPZ.
Table \ref{tab:medians} reports the median values of \mstar, \logoh, and sSFR
for the MEGA dataset for 6 mass bins within the 7 redshift bins shown in Fig. \ref{fig:redshift}.

%
\begin{table*} 
\setlength{\tabcolsep}{3pt}
\caption{Median stellar masses, (PP04N2) O/H, and sSFR in MEGA redshift bins$^{\mathrm a}$}
{
\begin{tabular}{llcccc}
\hline 
\multicolumn{1}{c}{Redshift bin} &
\multicolumn{1}{c}{Mass bin} &
\multicolumn{1}{c}{Number} &
\multicolumn{1}{c}{Log(\mstar)} &
\multicolumn{1}{c}{\logoh} &
\multicolumn{1}{c}{Log(sSFR)} \\
&&&
\multicolumn{1}{c}{(\msun)} &
\multicolumn{1}{c}{(PP04N2)} &
\multicolumn{1}{c}{(\msunyr)} \\
\hline
$z \leq\ 0.1$          & $\log({\mathrm M}_*)<8.5$         & 128 &   7.76$^{+0.31}_{-0.58}$ &   7.92$^{+0.12}_{-0.22}$ &  -9.61$^{+0.28}_{-0.23}$\\ 
 & $8.5\leq \log({\mathrm M}_*) < 9$ &  24 &   8.69$^{+0.11}_{-0.09}$ &   8.16$^{+0.10}_{-0.10}$ &  -9.78$^{+0.60}_{-0.18}$\\ 
 & $9\leq\ \log({\mathrm M}_*) < 9.5$ &  34 &   9.18$^{+0.12}_{-0.11}$ &   8.39$^{+0.09}_{-0.13}$ &  -9.84$^{+0.46}_{-0.18}$\\ 
 & $9.5\leq\ \log({\mathrm M}_*) < 10$ &  18 &   9.64$^{+0.15}_{-0.05}$ &   8.39$^{+0.19}_{-0.21}$ &  -9.94$^{+0.30}_{-0.11}$\\ 
 & $10\leq\ \log({\mathrm M}_*) < 10.5$ &  26 &  10.28$^{+0.12}_{-0.13}$ &   8.69$^{+0.08}_{-0.14}$ &  -9.99$^{+0.16}_{-0.15}$\\ 
 & $\log({\mathrm M}_*)>10.5$        &  27 &  10.71$^{+0.12}_{-0.11}$ &   8.75$^{+0.04}_{-0.03}$ & -10.36$^{+0.30}_{-0.62}$\\ 
\\ 
$0.1 < z \leq\ 0.4$    & $\log({\mathrm M}_*)<8.5$         &   1 &   8.09&   8.16 &  -8.91\\ 
 & $8.5\leq \log({\mathrm M}_*) < 9$ &  16 &   8.83$^{+0.08}_{-0.16}$ &   8.22$^{+0.08}_{-0.04}$ &  -9.22$^{+0.11}_{-0.32}$\\ 
 & $9\leq\ \log({\mathrm M}_*) < 9.5$ &  39 &   9.35$^{+0.09}_{-0.23}$ &   8.34$^{+0.10}_{-0.07}$ &  -9.52$^{+0.27}_{-0.33}$\\ 
 & $9.5\leq\ \log({\mathrm M}_*) < 10$ &  84 &   9.71$^{+0.12}_{-0.11}$ &   8.53$^{+0.11}_{-0.10}$ &  -9.69$^{+0.30}_{-0.24}$\\ 
 & $10\leq\ \log({\mathrm M}_*) < 10.5$ &  18 &  10.11$^{+0.09}_{-0.05}$ &   8.62$^{+0.06}_{-0.11}$ &  -9.78$^{+0.25}_{-0.24}$\\ 
 & $\log({\mathrm M}_*)>10.5$        &   6 &  10.57$^{+0.10}_{-0.04}$ &   8.60$^{+0.04}_{-0.09}$ &  -9.72$^{+0.08}_{-0.05}$\\ 
\\ 
$0.4 < z \leq\ 0.7$    & $\log({\mathrm M}_*)<8.5$         &   9 &   8.38$^{+0.02}_{-0.16}$ &   8.19$^{+0.04}_{-0.03}$ &  -8.58$^{+0.30}_{-0.36}$\\ 
 & $8.5\leq \log({\mathrm M}_*) < 9$ &  12 &   8.82$^{+0.15}_{-0.13}$ &   8.25$^{+0.05}_{-0.05}$ &  -9.02$^{+0.36}_{-0.17}$\\ 
 & $9\leq\ \log({\mathrm M}_*) < 9.5$ &  49 &   9.34$^{+0.09}_{-0.07}$ &   8.35$^{+0.13}_{-0.07}$ &  -9.45$^{+0.25}_{-0.14}$\\ 
 & $9.5\leq\ \log({\mathrm M}_*) < 10$ &  84 &   9.69$^{+0.09}_{-0.08}$ &   8.44$^{+0.09}_{-0.08}$ &  -9.41$^{+0.18}_{-0.21}$\\ 
 & $10\leq\ \log({\mathrm M}_*) < 10.5$ &  12 &  10.12$^{+0.10}_{-0.04}$ &   8.59$^{+0.02}_{-0.11}$ &  -9.37$^{+0.22}_{-0.12}$\\ 
 & $\log({\mathrm M}_*)>10.5$        &   3 &  10.53$^{+0.03}_{-0.00}$ &   8.64$^{+0.05}_{-0.01}$ &  -9.57$^{+0.15}_{-0.05}$\\ 
\\ 
$0.7 < z \leq\ 0.9$    & $\log({\mathrm M}_*)<8.5$         &   6 &   8.10$^{+0.18}_{-0.18}$ &   8.04$^{+0.12}_{-0.03}$ &  -7.58$^{+0.46}_{-0.26}$\\ 
 & $8.5\leq \log({\mathrm M}_*) < 9$ &  16 &   8.73$^{+0.12}_{-0.05}$ &   8.17$^{+0.18}_{-0.20}$ &  -8.07$^{+0.39}_{-0.38}$\\ 
 & $9\leq\ \log({\mathrm M}_*) < 9.5$ &  45 &   9.34$^{+0.11}_{-0.07}$ &   8.33$^{+0.11}_{-0.07}$ &  -9.13$^{+0.43}_{-0.28}$\\ 
 & $9.5\leq\ \log({\mathrm M}_*) < 10$ &  91 &   9.73$^{+0.08}_{-0.11}$ &   8.44$^{+0.11}_{-0.09}$ &  -9.21$^{+0.27}_{-0.29}$\\ 
 & $10\leq\ \log({\mathrm M}_*) < 10.5$ &  22 &  10.21$^{+0.08}_{-0.12}$ &   8.60$^{+0.07}_{-0.08}$ &  -9.23$^{+0.27}_{-0.24}$\\ 
 & $\log({\mathrm M}_*)>10.5$        &  20 &  10.74$^{+0.24}_{-0.08}$ &   8.69$^{+0.04}_{-0.05}$ &  -9.54$^{+0.34}_{-0.26}$\\ 
\\ 
$0.9 < z \leq\ 1.8$    & $\log({\mathrm M}_*)<8.5$         &   0 & \multicolumn{1}{c}{$-$} &   \multicolumn{1}{c}{$-$} &   \multicolumn{1}{c}{$-$}\\ 
 & $8.5\leq \log({\mathrm M}_*) < 9$ &   1 &   8.74&   7.49 &  -7.98\\ 
 & $9\leq\ \log({\mathrm M}_*) < 9.5$ &   2 &   9.20$^{+0.02}_{-0.02}$ &   8.08$^{+0.05}_{-0.05}$ &  -8.88$^{+0.08}_{-0.08}$\\ 
 & $9.5\leq\ \log({\mathrm M}_*) < 10$ &  14 &   9.81$^{+0.11}_{-0.17}$ &   8.41$^{+0.04}_{-0.06}$ &  -8.81$^{+0.28}_{-0.17}$\\ 
 & $10\leq\ \log({\mathrm M}_*) < 10.5$ &  14 &  10.24$^{+0.05}_{-0.11}$ &   8.52$^{+0.02}_{-0.03}$ &  -8.67$^{+0.06}_{-1.06}$\\ 
 & $\log({\mathrm M}_*)>10.5$        &  12 &  10.68$^{+0.17}_{-0.13}$ &   8.54$^{+0.07}_{-0.03}$ &  -9.12$^{+0.31}_{-0.81}$\\ 
\\ 
$1.8 < z \leq\ 2.8$    & $\log({\mathrm M}_*)<8.5$         &   0 & \multicolumn{1}{c}{$-$} &   \multicolumn{1}{c}{$-$} &   \multicolumn{1}{c}{$-$}\\ 
 & $8.5\leq \log({\mathrm M}_*) < 9$ &   2 &   8.74$^{+0.07}_{-0.07}$ &   8.12$^{+0.01}_{-0.01}$ &  -7.22$^{+0.07}_{-0.07}$\\ 
 & $9\leq\ \log({\mathrm M}_*) < 9.5$ &   5 &   9.28$^{+0.03}_{-0.10}$ &   8.25$^{+0.03}_{-0.05}$ &  -7.69$^{+0.05}_{-0.10}$\\ 
 & $9.5\leq\ \log({\mathrm M}_*) < 10$ &  33 &   9.69$^{+0.16}_{-0.10}$ &   8.24$^{+0.08}_{-0.02}$ &  -8.17$^{+0.16}_{-0.30}$\\ 
 & $10\leq\ \log({\mathrm M}_*) < 10.5$ &  24 &  10.18$^{+0.12}_{-0.07}$ &   8.28$^{+0.09}_{-0.07}$ &  -8.76$^{+0.26}_{-0.26}$\\ 
 & $\log({\mathrm M}_*)>10.5$        &  24 &  10.89$^{+0.25}_{-0.26}$ &   8.46$^{+0.03}_{-0.14}$ &  -9.21$^{+0.20}_{-0.12}$\\ 
\\ 
$2.8 < z \leq\ 3.8$    & $\log({\mathrm M}_*)<8.5$         &   2 &   8.41$^{+0.02}_{-0.02}$ &   7.93$^{+0.01}_{-0.01}$ &  -7.90$^{+0.26}_{-0.26}$\\ 
 & $8.5\leq \log({\mathrm M}_*) < 9$ &   6 &   8.82$^{+0.12}_{-0.05}$ &   7.88$^{+0.22}_{-0.06}$ &  -7.61$^{+0.11}_{-0.16}$\\ 
 & $9\leq\ \log({\mathrm M}_*) < 9.5$ &  12 &   9.35$^{+0.07}_{-0.06}$ &   8.12$^{+0.04}_{-0.17}$ &  -8.09$^{+0.23}_{-0.20}$\\ 
 & $9.5\leq\ \log({\mathrm M}_*) < 10$ &  24 &   9.71$^{+0.14}_{-0.05}$ &   8.16$^{+0.01}_{-0.15}$ &  -8.19$^{+0.18}_{-0.42}$\\ 
 & $10\leq\ \log({\mathrm M}_*) < 10.5$ &  16 &  10.12$^{+0.18}_{-0.05}$ &   8.21$^{+0.05}_{-0.04}$ &  -8.77$^{+0.29}_{-0.19}$\\ 
 & $\log({\mathrm M}_*)>10.5$        &   8 &  10.75$^{+0.06}_{-0.14}$ &   8.19$^{+0.09}_{-0.02}$ &  -8.86$^{+0.38}_{-0.31}$\\ 
\\ 
\hline
\label{tab:medians} 
\end{tabular} 
}
\vspace{-\baselineskip}
\begin{flushleft}
$^{\mathrm a}$~Medians of values within each redshift bin;
the upper and lower values correspond to the 75\% and 25\% quantile levels,
respectively. We have not considered the AMAZE galaxy, LnA1689$-$2, at $z\,=\,4.87$.
\end{flushleft}
\end{table*}

\subsection{Redshift variation of O/H and sSFR}
\label{sec:redshiftvariation}

\begin{figure*}
\vspace{\baselineskip}
\hbox{
\includegraphics[width=0.475\textwidth]{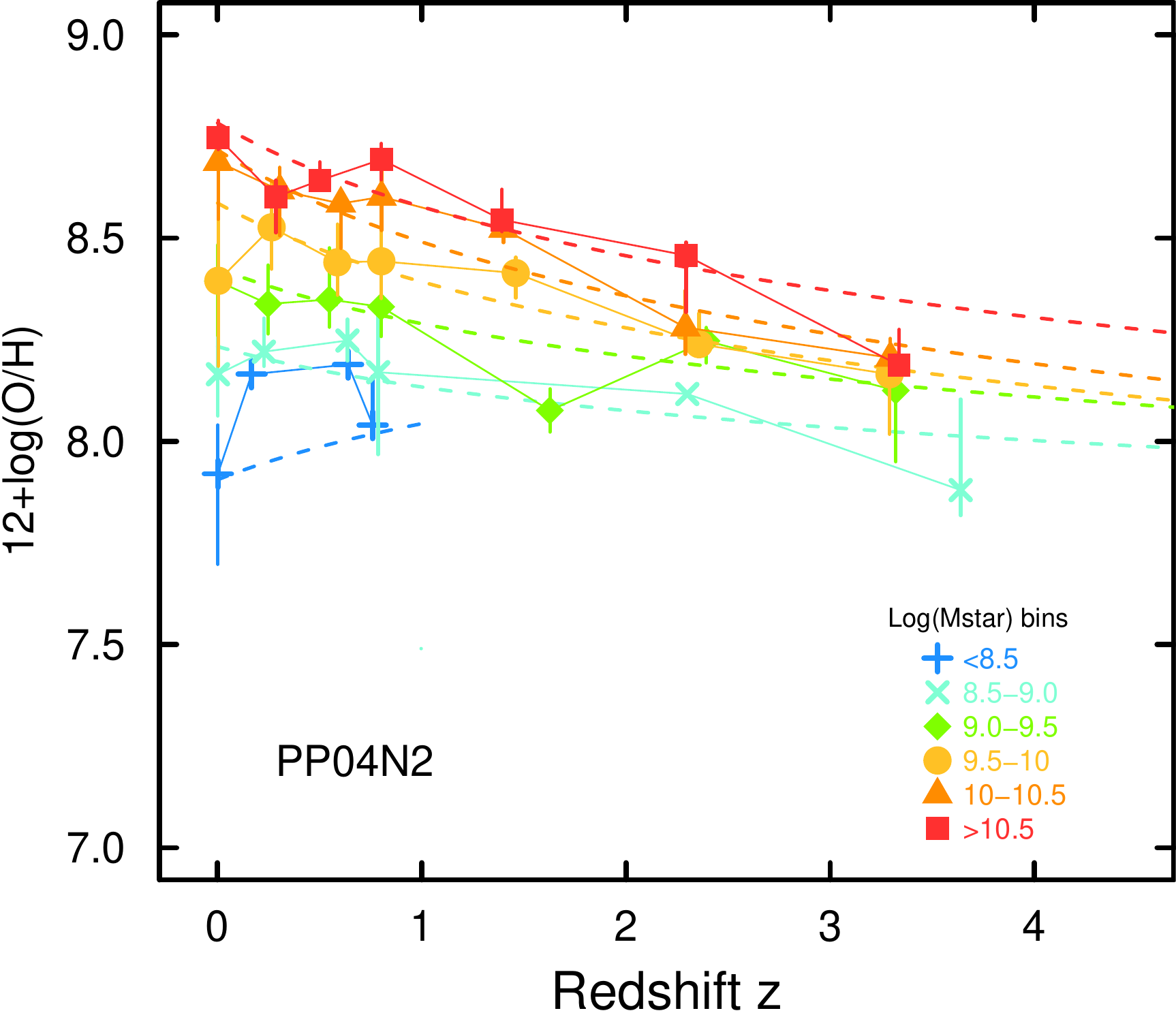}
\hspace{0.3cm}
\includegraphics[width=0.475\textwidth]{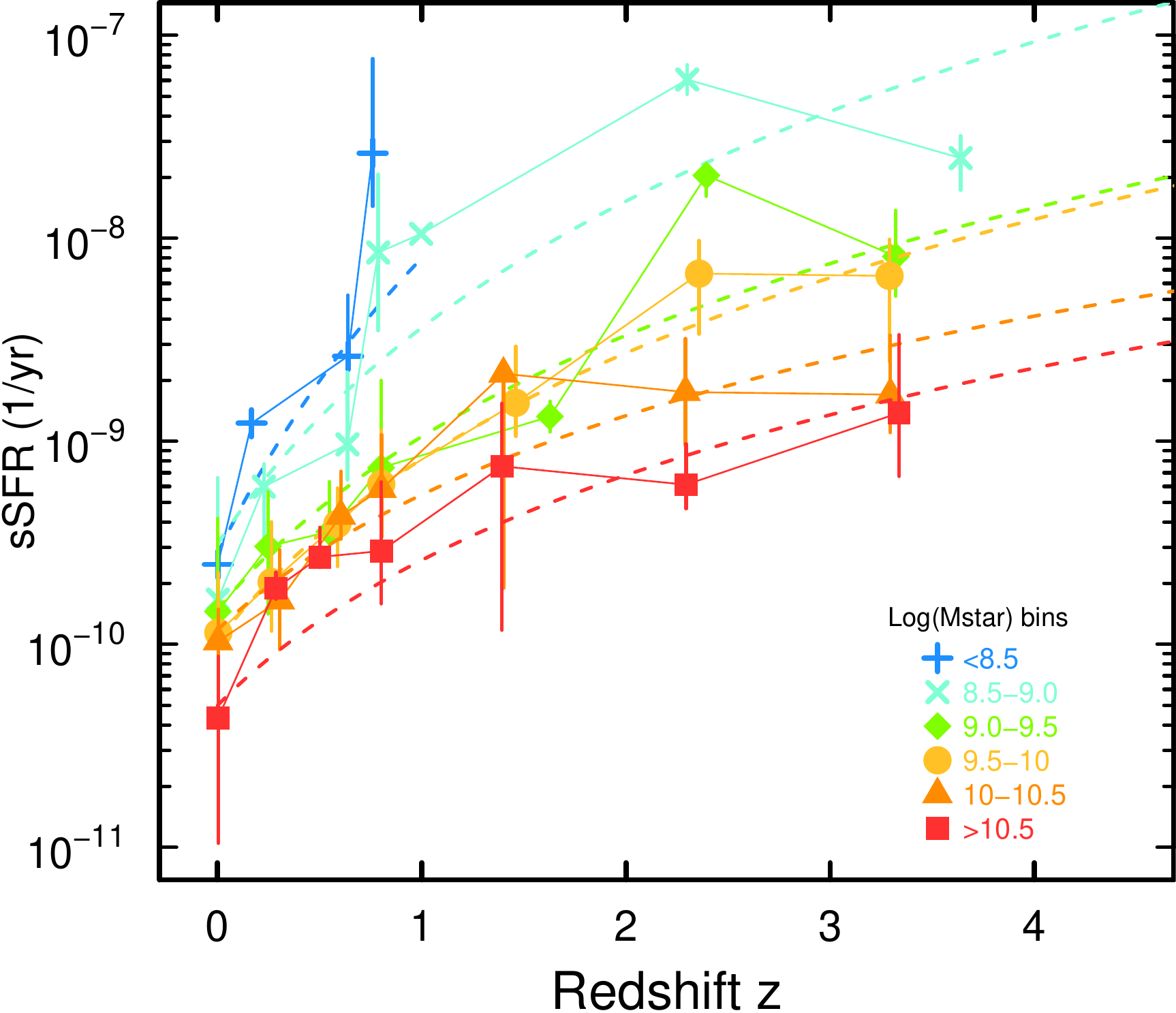}
}
\caption{Binned measurements of \logoh\ and sSFR as a function of redshift
using the MEGA dataset. 
As in previous figures, the O/H calibration is PP04N2;
the mass bins are shown in the lower right-hand corner.
Error bars correspond to the 25th percentile of the vertical
parameter within each mass bin.
The curves are an approximation to the observed
trends obtained by adopting the formulation of \citet{karim11} based on
separable functions of \mstar\ and $z$: sSFR, O/H $\propto\ M_*^\beta\ (1+z)^n$.
The mean power-law indices, $n$, averaged over all mass bins (except the lowest one) are: 
\meann({\rm O/H})\,=\,$-0.57\,\pm\,0.17$ for \logoh\,$\propto (1+z)^n$ (left panel), and 
\meann({\rm sSFR})\,=\,$2.79\,\pm\,0.52$ for sSFR\,$\propto (1+z)^n$ (right). 
}
\label{fig:ohssfrvsz}
\end{figure*}

The trends with redshift of the MEGA dataset are shown in
Fig. \ref{fig:ohssfrvsz} where \logoh\ and sSFR are plotted vs. redshift $z$.  
Following \citet{karim11}, we have fit the redshift variation with
separable functions in \mstar\ and $z$: 

$$
\mathrm{sSFR}, \mathrm{O/H} ({\mathrm M}_*,z) \propto {\mathrm M}_*^\beta\ (1+z)^n.
$$

\noindent
The dependence on stellar mass is encompassed in the power-law index $\beta$, 
and the $z$ dependence in the power-law index (slope in log space), $n$. 
The symbols in Fig. \ref{fig:ohssfrvsz} correspond to data binned in redshift and in stellar mass as given in
the legend (see Fig. \ref{fig:redshift} for redshift intervals); 
the error bars give the 25\% quantiles of the data within each bin.

For the trend of sSFR\,$\propto (1+z)^{n({\rm sSFR})}$, the mean power-law index, \meann(sSFR), 
averaged over all mass bins (except the lowest one,
because of the lack of low-mass galaxies at $z \ga 1$) is: 
\meann(sSFR)\,=\,$2.8\,\pm\,0.5$. 
For the \mstar\ bins between dex(8.5) and dex(10)\,\msun,
$n$ is relatively constant: $3.1\,\pm\,0.37$.
This value is roughly consistent with $n({\rm sSFR})\sim 3.4-3.5$ 
to $z \sim 2$ as reported by \citet{oliver10,karim11}. 
For larger \mstar\ (\mstar\,$\geq10^{10}$\,\msun), the index decreases to $n({\rm sSFR})\,=\,2.2-2.4$.
Similar slopes and such a flattening are also seen in the highly star-forming sub-sample of COSMOS galaxies
described by \citet{karim11}; the MEGA dataset probably represents a similarly highly star-forming
sample, at least at the higher redshifts. 

The redshift variation of O/H is represented by
the mean power-law index \meann({\rm O/H}), averaged over all mass bins (except
the lowest one as above): 
\meann({\rm O/H})\,=\,$-0.57\,\pm\,0.17$ for \logoh\,$\propto (1+z)^{n({\rm O/H})}$ (left panel of Fig. \ref{fig:ohssfrvsz}).
From the relatively small standard deviation, and visually evident in Fig. \ref{fig:ohssfrvsz}, 
it is apparent that the index $n({\rm O/H})$ is much more constant over variations in \mstar\ than 
the equivalent index $n({\rm sSFR})$ for sSFR.
Table \ref{tab:ohvszbins} gives the fitted coefficients for the (PP04N2) metallicity redshift variation
of the MEGA dataset;
these are the equations describing the dashed curves in the left panel of Fig. \ref{fig:ohssfrvsz} 
and subsequent figures.

\begin{table} 
\setlength{\tabcolsep}{3pt}
\caption {Fitted (PP04N2) O/H redshift variation of the MEGA dataset$^{\mathrm a}$}
{\scriptsize
\begin{tabular}{crcccc}
\hline 
\multicolumn{1}{c}{Mass bin} &
\multicolumn{1}{c}{Number} &
\multicolumn{1}{c}{$a$} &
\multicolumn{1}{c}{$b$} &
\multicolumn{1}{c}{Median} &
\multicolumn{1}{c}{RMS} \\
\multicolumn{1}{c}{log(\msun)} & \multicolumn{1}{c}{redshift} & & &
\multicolumn{1}{c}{residual} &
\multicolumn{1}{c}{$\sigma$} \\
& \multicolumn{1}{c}{points} \\
\hline
$8.5-9.0$   &  77 & $8.23\,\pm\,0.03$ & $-0.33\,\pm\,0.11$ & ~$0.010$ & $0.18$\\
$9.0-9.5$   & 186 & $8.43\,\pm\,0.02$ & $-0.45\,\pm\,0.07$ & ~$0.002$ & $0.15$\\
$9.5-10.0$  & 349 & $8.59\,\pm\,0.01$ & $-0.64\,\pm\,0.05$ & $-0.002$ & $0.14$\\
$10.5-10.5$ & 132 & $8.72\,\pm\,0.02$ & $-0.75\,\pm\,0.05$ & ~$0.012$ & $0.13$\\
$>10.5$     & 100 & $8.78\,\pm\,0.02$ & $-0.68\,\pm\,0.06$ & ~$0.005$ & $0.14$\\
\hline
\label{tab:ohvszbins} 
\end{tabular} 
}
\vspace{-\baselineskip}
\begin{flushleft}
$^{\mathrm a}$~Coefficients for robust fits to the individual data 
points (within each mass bin) of the
form \logoh\,=\,$a$\ + $b\,\log(1+z)$; $b$ corresponds to $n$(O/H) as
described in the text.\\
\end{flushleft}
\end{table}

\begin{figure*}
\vspace{\baselineskip}
\hbox{
\includegraphics[width=0.475\textwidth]{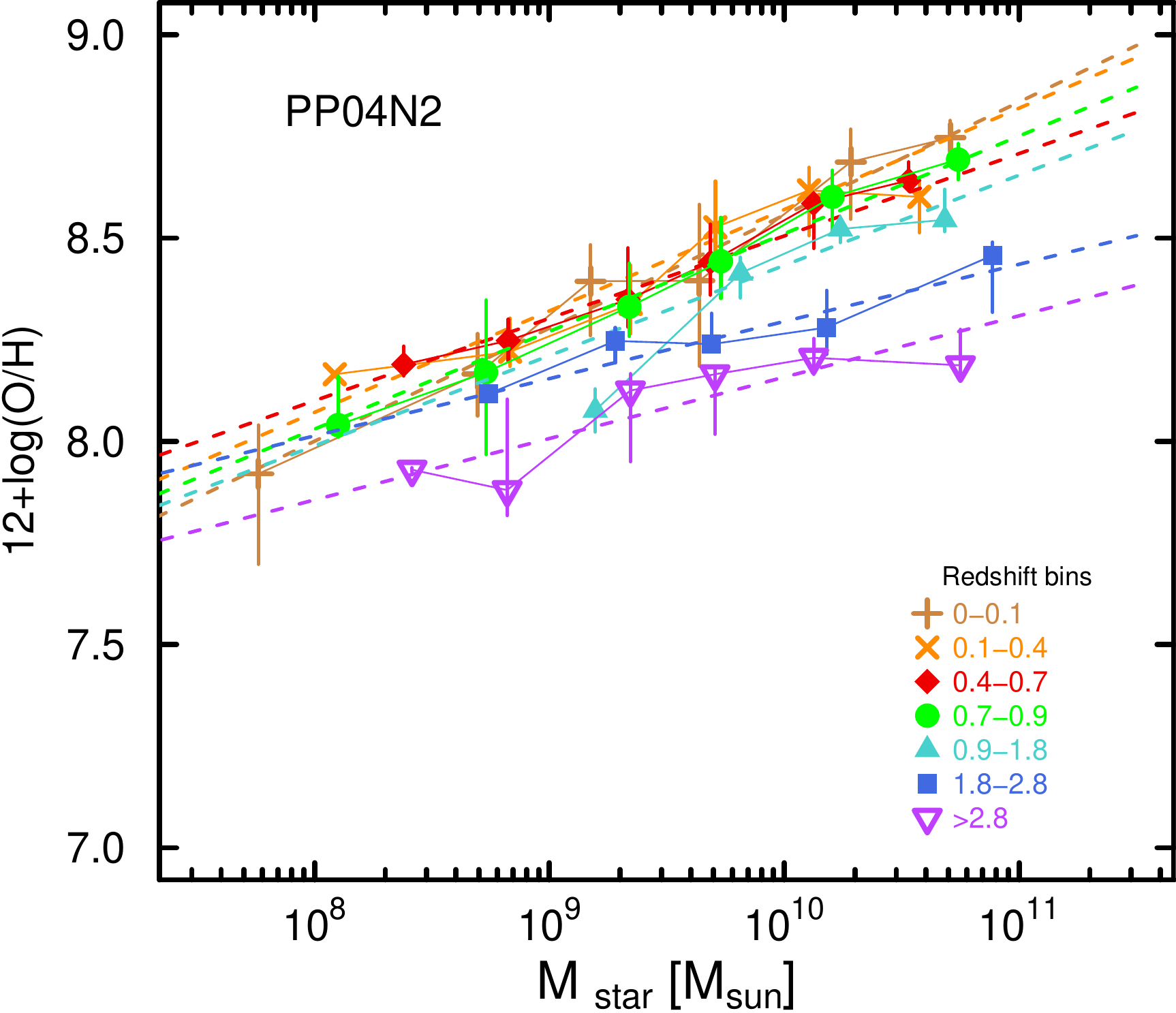}
\hspace{0.3cm}
\includegraphics[width=0.475\textwidth]{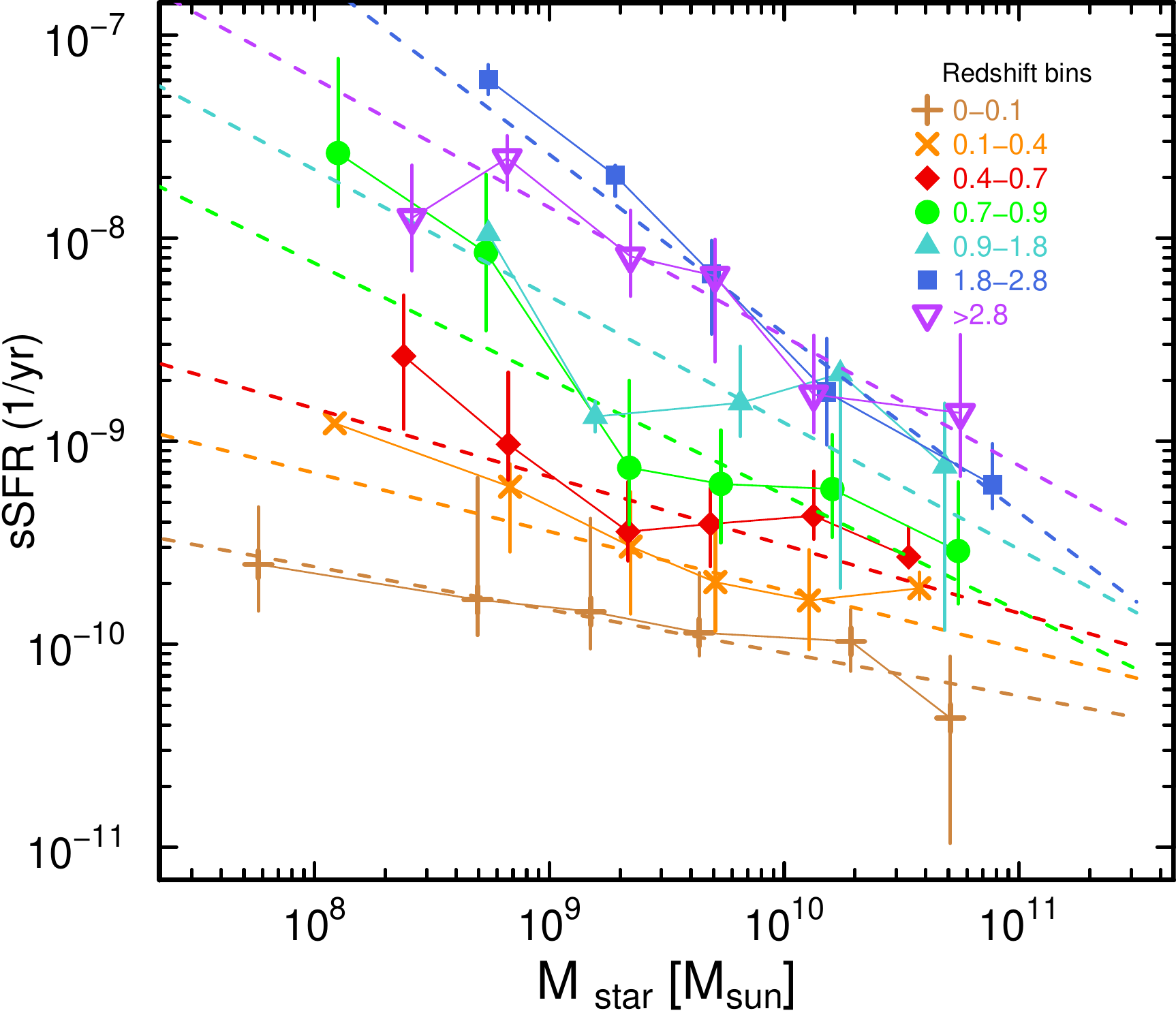}
}
\caption{Binned measurements of \logoh\ and sSFR as a function of \mstar\
using the MEGA dataset. 
As in previous figures, the representative O/H calibration is PP04N2;
the redshift bins are shown in the corners of the figures (see also Fig. \ref{fig:redshift}). 
Error bars correspond to the 25th percentile of the vertical
parameter within each redshift bin.
As in Fig. \ref{fig:ohssfrvsz}, the curves are an approximation to the observed
trends obtained by adopting the formulation of \citet{karim11} based on
separable functions with the redshift trend given by $(1+z)^n$, and the
\mstar\ variation by $M_*^\beta$.
The mean power-law indices, $\beta$, averaged over all redshift bins are: 
\meanb({\rm O/H})\,=\,$0.21\,\pm\,0.05$ for \logoh\,$\propto$\,\mstar$^{\beta({\rm O/H})}$ (left panel), and 
\meanb({\rm sSFR})\,=\,$-0.51\,\pm\,0.24$ for sSFR\,$\propto$\,\mstar$^{\beta({\rm sSFR})}$ (right). 
}
\label{fig:ohssfrvsmstar}
\end{figure*}

\subsection{Variation of O/H and sSFR with stellar mass}
\label{sec:massvariation}

We now examine the \mstar\ dependence of the redshift variations of O/H and sSFR.
Fig. \ref{fig:ohssfrvsmstar} shows \logoh\ vs. \mstar\ (left panel) and
sSFR vs. \mstar\ (right), binned into different redshift bins;
these are equivalent to the changes with redshift of the MZR (left panel) and the SFMS (right).

Here we assess $\beta$ in the separated formalism as above:
sSFR, O/H $\propto M_*^\beta\ (1+z)^n$.
$\beta$(O/H) corresponds roughly to the slope (power-law index) of the MZR,
and $\beta$(sSFR) to the slope of the SFMS.
For the \mstar\ trend of O/H, \meanb\ (averaged over all mass bins) is:
\meanb(O/H)\,=\,$0.21\,\pm\,0.05$.
Both the normalization and the slope of the MZR are rather constant up to $z \sim 2$:
$\beta\,=\,0.24\,\pm\,0.03$, 
\logoh(\mstar\,=\,dex(9)\,\msun)\,=\,$8.28\,\pm\,0.04$
(averages and standard deviations over 5 equally-weighted redshift bins for $z < 1.8$).
However, for $z > 2$, both the normalization and the slope gradually decrease:
$\beta\,=\,0.15\,\pm\,0.03$ and \logoh(\mstar\,=\,dex(9)\,\msun)\,=\,$8.01\,\pm\,0.03$
within the highest ($z \sim 3$) redshift bin.

The slope $\beta$(O/H) for the MEGA dataset at $z \sim 1.4$ of $0.23\,\pm\,0.02$ is slightly
steeper than that by \citet{yabe14} who find $\beta$\,=\,0.15,
but shallower than the slope of $\sim 0.3$ found by \citet{liu08} using PP04N2 in a similar
redshift range (both samples are also included in the MEGA dataset).
The steeper slopes we find relative to \citet{yabe14} can be attributed to their use
of a Salpeter IMF, rather than the \citet{chabrier03} IMF used here.
In the redshift range $z \sim 0.5-0.9$, \citet{cowie08} find an MZR slope
of $0.13-0.17$ using the KK04 and T04 O/H calibrations and a Salpeter IMF;
this is also somewhat shallower than the PP04N2 $\beta$\,=\,$0.24\,\pm\,0.02$ in a similar redshift range,
but with the Chabrier IMF.
The steeper slopes we find for the MEGA dataset are consistent with those
found by \citet{zahid11} for DEEP2 galaxies at $z \sim 0.8$ using the KK04 O/H calibration
(and a Chabrier IMF).
As discussed by \citet{zahid11}, differences in fitting procedures are an 
important consideration in comparing slopes of the MZR,
but the O/H calibration is also important.
\citet{kewley08} illustrate that both the slope (at the low-mass end) and
the absolute O/H determination depend strongly on the calibration. 
Thus, the consistency of the MZR slopes $\beta$ relative to previous
work lends confidence our approach.

For the trends of sSFR with \mstar, corresponding to the SFMS,
averaging over all redshifts gives \meanb(sSFR)\,=\,$-0.51\,\pm\,0.24$
(average and standard deviation over the individual redshift bins);
the mean slope is poorly determined because of the (possibly spurious, see below)
steepening toward high $z$.
At $z \simeq 0$, 
we find $\beta$(sSFR)$\,=\,-0.21\,\pm\,0.025$\footnote{This slope is derived
from all data at $z \simeq 0$, while the slope of $-0.19\,\pm\,0.02$ in Sect. \ref{sec:scaling}
is found from the LVL$+$KINGFISH galaxies only; the two slopes
are in good agreement.}.
At $z\sim0.25$, we find a steeper slope, $\beta$(sSFR)$\,=\,-0.29\,\pm\,0.07$, roughly consistent
with the sSFR vs. \mstar\ power-law index of $\sim -0.4$ estimated by \citet{karim11} and by \citet{speagle14}
for $z \approx 0.3$.
At $z \sim 3$, $\beta$(sSFR)$\,=\,-0.64\,\pm\,0.09$; 
the observed steepening of $\beta$(sSFR) toward
higher redshift evident in Fig. \ref{fig:ohssfrvsmstar} is inconsistent with
the results of \citet{speagle14} who
find steeper slopes with increasing redshift in (log) SFR vs. \mstar, corresponding to  
shallower slopes in (log) sSFR.

Indeed, the MEGA dataset does not show clear evidence for a SFMS within
individual redshift bins (see Fig. \ref{fig:ms}); this could be because the galaxies at higher redshift are selected
basically for a constant SFR (see Table \ref{tab:samples}) rather than
a selection based on \mstar.
Such a selection can result in a basically flat trend of SFR with \mstar\
\citep[e.g.,][]{erb06b,lee13,renzini15},
which produces a steep dependence of sSFR with \mstar, sSFR$\,\propto$\,\mstar$^{-1}$,
similar to the behavior of the MEGA dataset at high $z$.
This is essentially a Malmquist bias since at low stellar masses, only
galaxies with relatively high SFR are selected.
Because the MEGA dataset requires emission lines in order to measure
metallicity spectroscopically, such an effect almost certainly
plays an important role \citep[e.g.,][]{juneau14}.
Ultimately, because of such selection effects,
the MEGA dataset may not be completely representative of the 
SFR-\mstar\ correlations at high redshift.
Nevertheless, it is the best dataset currently available for assessing
the evolution of the MZR.

\subsection{Redshift invariance of the FPZ?}
\label{sec:fpzinvariance}

One way to assess
the redshift invariance of the FPZ is by comparing the coefficients
for redshift variation discussed above in Sect. \ref{sec:redshiftvariation}
with those for the FPZ.
We thus performed multi-variable linear regressions on the MEGA dataset 
for \logoh\ as a function of \mstar\ and redshift, and the same for sSFR.
Performing a robust fit\footnote{For all statistical calculations we use R,
a free software environment for statistical computing and graphics,
{\it https://www.r-project.org/}.},
we find (for the PP04N2 calibration):

\begin{equation}
12+\log({\rm O/H})\,=\,0.27\,\log({\rm M}_*) - 0.59\,\log(1+z) + 5.89
\label{eqn:ohvsz}
\end{equation}

\noindent
with a residual standard error of $\sim$0.15\,dex,
and

\begin{equation}
\log({\rm sSFR})\,=\,-0.29\,\log({\rm M}_*) + 2.88\,\log(1+z) - 7.16 
\label{eqn:ssfrvsz}
\end{equation}

\noindent
with a residual standard error of $\sim$0.44\,dex,
The FPZ in Eqn. (\ref{eqn:fpzall}) can be expressed as a function of sSFR,
rather than SFR:
\begin{align}
12+\log({\rm O/H}) & = -0.14\,\log {\rm (sSFR)} + 0.23\,\log ({\rm M_*)} + 4.82  
\label{eqn:fpzallssfr}
\end{align}


By inserting the redshift variation of sSFR given by Eqn. (\ref{eqn:ssfrvsz}) in 
Eqn. (\ref{eqn:fpzallssfr}) for the FPZ, and comparing it with the redshift
variation of O/H given by Eqn. (\ref{eqn:ohvsz}), we can
compare the resulting difference equation term by term and assess the redshift invariance of the FPZ;
we would expect $\approx 0$ in such a case.
The resulting coefficient for the difference (FPZ $-$ $z$ fits) of the log(\mstar) term is $0.0004$, consistent with 0.
For log($1+z$), we find a difference of $0.178$, roughly consistent with 0 to within the residual standard errors of the fits.
The resulting difference for the constant term is $-0.051$, again consistent with 0 within the standard errors.
Although this result pertains to PP04N2, similarly small difference coefficients are
obtained for the D02 and PP04O3N2 calibrations.
\textit{We thus conclude that the FPZ is approximately redshift invariant to within 0.15$-$0.16\,dex}
(see Sect. \ref{sec:fp}); for typical galaxy populations
the increase of sSFR with redshift is compensated by the decrease in O/H \citep[although see][]{wuyts14}.

\begin{figure}
\vspace{\baselineskip}
\hbox{
\includegraphics[width=0.47\textwidth]{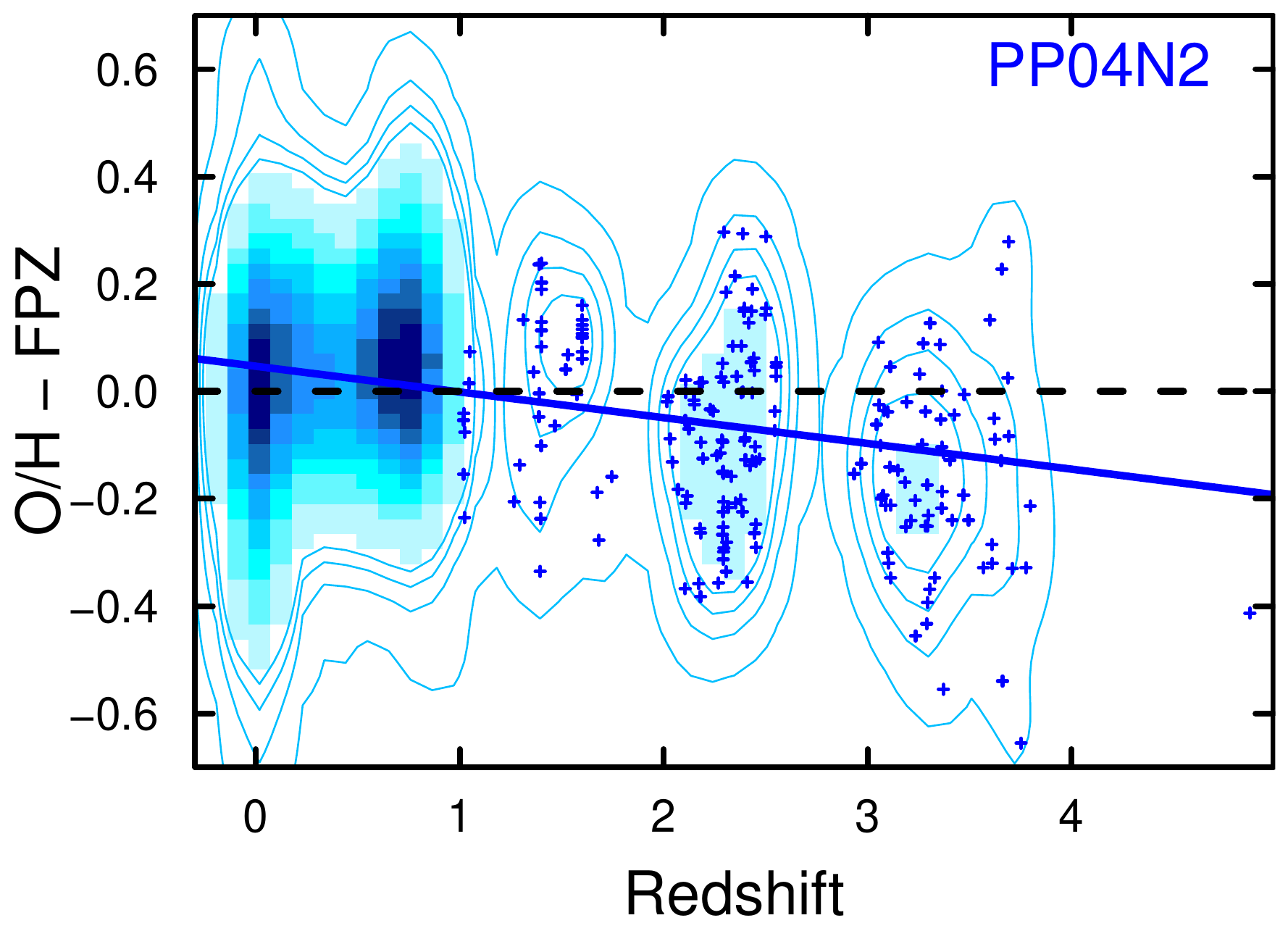}
}
\caption{FPZ residuals for the MEGA dataset with the PP04N2 calibration
as a function of redshift.
The color scale corresponds to density of data points, and the contours show
the full dataset.
Individual points are also plotted for $z\geq1$.
The horizontal dashed line guides the eye for zero residuals, while
the solid line shows the best-fit robust regression given by Eqn. \ref{eqn:fpzresvsz}.
}
\label{fig:fpresidualsvsz}
\end{figure}

Nevertheless, the log($1+z$) difference coefficient of $\sim 0.18$\,dex is slightly larger
than would be expected given the residuals of the (PP04N2) FPZ, $\sim 0.16$\,dex.
This implies that the FPZ is not a perfect formulation of the redshift evolution of metallicity.
Therefore, to investigate the amplitude of the residual trend with redshift, in
Fig. \ref{fig:fpresidualsvsz} we have plotted the FPZ O/H residuals vs. redshift.
The trend (shown by the solid line) is significant with:

\begin{align}
12+\log({\rm O/H}) - {\rm FPZ(PP04N2)} & = \nonumber \\
(-0.048\,\pm\,0.006)\,z + 0.047\,\pm\,0.007
\label{eqn:fpzresvsz}
\end{align}

\noindent
This implies that at $z\sim 3.5$, the FPZ predicts metallicities \logoh\ that are
roughly $0.17$\,dex too large.
However, the residual standard error of the (PP04N2) fit in Eqn. (\ref{eqn:fpzresvsz}) is $0.16$\,dex,
equivalent to the residuals of the FPZ itself;
indeed, the spread of residuals at $z\sim0$ is as large or larger than the spread of residuals
at $z\ga3$.
Thus, while the current data suggest that the FPZ may not fully describe metallicity
evolution (or its lack thereof related to SFR), the discrepancies are within
the overall noise in the estimation.
The data at $z\ga 3$ are still relatively sparse, however, and more data 
with accurate metallicity measurements should help in confirming (or refuting)
this conclusion.

\section{Discussion and summary}
\label{sec:discussion}

The FPZ presented in Sect. \ref{sec:fp} is based on the
hypothesis that the curvature in the MZR at high stellar masses is compensated by the inflection
in the SFMS.
Our results show that this hypothesis is reasonably good, at least to within $\sim$0.16\,dex in \logoh.
Comparison of the FPZ with the FMR shows that both formulations adequately represent
the mutual correlations of \mstar, SFR, and O/H, but
also that the O/H calibration is crucial;
applying the FMR to arbitrary samples can result in metallicity offsets as large as $-0.2$ to $-0.3$\,dex,
compared with the three O/H calibrations considered here.

Some groups have concluded that at a given redshift
or over a narrow range of redshifts, that the MZR does not depend on SFR
\citep[e.g.,][]{wuyts14,sanders15}.
However,
the parameters of the MEGA sample in specific redshift intervals, given in Table \ref{tab:medians}, are
comparable to the other samples used to draw these conclusions. 
It is likely that the broad parameter space spanned by the MEGA dataset
contribute to the differences in the outcome.
The PCA analysis presented here could also play a role;
indeed, if we fit the MEGA data with a simple multi-variable regression 
of \logoh\ with respect to
\mstar, SFR, and redshift, we find very little dependence of O/H on SFR.
This is because of the strong dependence of SFR (and sSFR) on redshift
through the increasing normalization of the SFMS (e.g., Sect. \ref{sec:coevo}). 
The mutual correlations of the variables underlying the FPZ must be
taken into account for any analysis considering the MZR and its dependence on SFR.

In conclusion, we have compiled a new MEGA dataset consisting of $\sim$1000 galaxies taken
from 19 individual samples spanning a wide range of stellar masses, SFRs, and
metallicities and covering redshifts from $z\simeq 0$ to $z\sim 3.7$. In
addition to larger numbers of high-$z$ galaxies, the main improvement of this
dataset over that of \citet{hunt12} is the common O/H calibrations derived for
the MEGA galaxies. The main results are as follows:

\noindent
\hangindent=0.05\linewidth
\hangafter=1
$\bullet$\ \ After examining the mutual correlations among these parameters,
a PCA of the MEGA dataset shows that the 3D parameter space can be described
by a plane, dubbed the FPZ.

\noindent
\hangindent=0.05\linewidth
\hangafter=1
$\bullet$\ \
The functional form of the (PP04N2) FPZ is given by $12+\log({\rm O/H})  = -0.14\,\log
{\rm (SFR)} + 0.37\,\log ({\rm M_*)} + 4.82$ over the entire mass and redshift range of the MEGA dataset. 

\noindent
\hangindent=0.05\linewidth
\hangafter=1
$\bullet$\ \
The mean O/H residuals of the FPZ over the MEGA dataset are 0.16\,dex
(for the PP04N2 calibration, slightly larger for D02 and PP04O3N2);
such residuals are 
smaller than those found previously, and consistent with trends found in
smaller galaxy samples with more limited ranges in \mstar, SFR, and O/H.

\noindent
\hangindent=0.05\linewidth
\hangafter=1
$\bullet$\ \
The FPZ is also found to be roughly invariant with redshift enabling an estimation of
metallicity accurate to within 0.16\,dex over roughly 5 orders
of magnitude in \mstar,
from $\ga 10^{6}$\,\msun\ to $\sim 10^{11}$\,\msun, 
up to $z\sim 3.7$.
An additional correction for redshift may be employed to increase 
slightly the accuracy of O/H estimates from the FPZ
for $z\ga 2$ (see Eqn. \ref{eqn:fpzresvsz}).


\section*{Acknowledgments}
We acknowledge the anonymous referee whose insightful comments greatly improved the paper.
We also thank G. Cresci for passing us the SDSS10 (KD02) data in electronic form, and are grateful to the DAVID network
({\it http://wiki.arcetri.astro.it/DAVID/WebHome}) for fostering a fruitful collaborative environment.
PD gladly acknowledges funding from the EU COFUND Rosalind Franklin program.

\label{lastpage}

\end{document}